\documentclass[aps,prb,twocolumn,showpacs]{revtex4}

\usepackage{graphicx}
\usepackage{amsmath}
\usepackage{amssymb}

\begin{document}

\title{ Enhancement of superconductivity by electronic nematicity in cuprate
superconductors}

\author{Zhangkai Cao$^{1}$, Yiqun Liu$^{2}$, Huaiming Guo$^{3}$, and Shiping Feng$^{1}$}
\email{spfeng@bnu.edu.cn}

\affiliation{$^{1}$Department of Physics, Beijing Normal University, Beijing 100875,
China}

\affiliation{$^{2}$School of Physics, Nanjing University, Nanjing 210093, China}

\affiliation{$^{3}$School of Physics, Beihang University, Beijing 100191, China}

\begin{abstract}
The cuprate superconductors are characterized by numerous ordering tendencies, with the
electronically nematic order being the most distinct form of order. Here the intertwinement of
the electronic nematicity with superconductivity in cuprate superconductors is studied based on
the kinetic-energy-driven superconductivity. It is shown that the optimized $T_{\rm c}$ takes a
dome-like shape with the weak and strong strength regions on each side of the optimal strength
of the electronic nematicity, where the optimized $T_{\rm c}$ reaches its maximum. This
dome-like shape nematic-order strength dependence of $T_{\rm c}$ thus indicates that the
electronic nematicity enhances superconductivity. Moreover, this nematic order induces the
anisotropy of the electron Fermi surface, where although the original electron Fermi surface
contour with the four-fold rotation symmetry is broken up into that with a residual two-fold
rotation symmetry, this electron Fermi surface contour with the two-fold rotation symmetry still
is truncated to form the disconnected Fermi arcs with the most spectral weight that locates at
around the tips of the Fermi arcs. Concomitantly, these tips of the Fermi arcs connected by the
scattering wave vectors ${\bf q}_{i}$ construct an octet scattering model, however, the partial
scattering wave vectors and their respective symmetry-corresponding partners occur with unequal
amplitudes, leading to these electronically ordered states being broken both rotation and
translation symmetries. As a natural consequence, the electronic structure is inequivalent
between the $k_{x}$ and $k_{y}$ directions in momentum space. These anisotropic features of the
electronic structure are also confirmed via the result of the autocorrelation of the
single-particle excitation spectra, where the breaking of the rotation symmetry is verified by
the inequivalence on the average of the electronic structure at the two Bragg scattering sites.
The theory also indicates that the order parameter of the electronic nematicity achieves its
maximum in the characteristic energy, and then decreases rapidly as the energy moves away from
the characteristic energy.
\end{abstract}


%\begin{keywords}
%\textbf{Superconducting transition temperature; Electronic nematicity; Rotation symmetry-breaking;
%Electronic ordered state; Cuprate superconductor}
%\end{keywords}

\maketitle

\section{Introduction}\label{Introduction}

The cuprate superconductors are structurally characterized by the stacking square-lattice
CuO$_{2}$ planes \cite{Bednorz86,Wu87}. At the half-filling, the charge degree of freedom in
the CuO$_{2}$ plane is quenched, and then the parent compound of cuprate superconductors is a
Mott insulator with an antiferromagnetic (AF) long-range order (AFLRO) \cite{Fujita12}. It is
widely believed that this Mott-insulating state occurs to be due to the very strong electron
correlation \cite{Anderson87}. However, when the charge carriers are introduced into the
CuO$_{2}$ plane, AFLRO disappears rapidly \cite{Fujita12} and soon thereafter is transformed
into a superconductor with the exceptionally high superconducting (SC) transition temperature
$T_{\rm c}$ \cite{Bednorz86,Wu87}. This marked transformation of the electronic states therefore
indicates that the strong electron correlation also plays an essential role in the appearance of
superconductivity \cite{Anderson87,Timusk99,Hufner08}. However, this strong electron correlation
also induces the system to find new way to lower its total energy, often by the spontaneous
breaking of the native symmetries of the square lattice underlying the CuO$_{2}$ plane
\cite{Vishik18,Comin16,Vojta09,Fradkin10,Fernandes19}. This is why associated with this
transformation is the coexistence and intertwinement of multiple symmetry-breaking orders and
superconductivity. In this case, the understanding of the interplay between a bewildering variety
of spontaneous symmetry-breaking orders and superconductivity is a central issue for cuprate
superconductors.

Among these spontaneous symmetry-breaking orders, the most distinct form of order is
electronically nematic order
\cite{Vojta09,Fradkin10,Fernandes19,Lawler10,Fujita14,Zheng17,Fujita19,Nakata18,Hinkov08,Sato17,Daou10,Taillefer15,Wang21,Ando02,Wu17},
which corresponds to that the electronic structure breaks the discrete rotation symmetry of the
square-lattice CuO$_{2}$ plane. By virtue of systematic studies using multiple
measurement techniques, a number of consequences from the electronic nematicity together with the
associated fluctuation phenomena have been identified experimentally in the well-defined regimes
of the phase diagrams as
\cite{Vojta09,Fradkin10,Fernandes19,Lawler10,Fujita14,Zheng17,Fujita19,Nakata18,Hinkov08,Sato17,Daou10,Taillefer15,Wang21,Ando02,Wu17}:
(i) the locally anisotropic single-particle behaviours and the related quasiparticle scattering
interference (QSI) \cite{Lawler10,Fujita14,Zheng17,Fujita19}, (ii) the globally anisotropic
single-particle features and the related distorted electron Fermi surface (EFS)\cite{Nakata18},
(iii) the anisotropic spin excitation spectra and the related magnetic susceptibility
\cite{Hinkov08,Sato17,Daou10,Taillefer15,Wang21}, and (iv) the transport anisotropy and the
related conductivity spectra \cite{Ando02,Wu17}. In particular, this nematic order has been
observed recently by Raman scattering \cite{Loret19}, while the Nernst effect and magnetic torque
measurements \cite{Sato17,Daou10,Taillefer15} show that the onset temperature of the
nematic-order signals in the Nernst coefficient and that in the magnetic susceptibility coincide
with the pseudogap crossover temperature $T^{*}$. More importantly, this nematic order has been
found experimentally
\cite{Vojta09,Fradkin10,Fernandes19,Lawler10,Fujita14,Zheng17,Fujita19,Nakata18,Hinkov08,Sato17,Daou10,Taillefer15,Wang21,Ando02,Wu17,Loret19}
to coexist with the spontaneous translation symmetry breaking such as charge order, and then
intertwines with superconductivity below $T_{\rm c}$. All these experimental observations
therefore offer strong evidences that the electronic nematicity is an integral part of the
essential physics of cuprate superconductors
\cite{Vojta09,Fradkin10,Fernandes19,Lawler10,Fujita14,Zheng17,Fujita19,Nakata18,Hinkov08,Sato17,Daou10,Taillefer15,Wang21,Ando02,Wu17,Loret19}.
However, in despite of these developments, the role of the electronic nematicity, such as whether
it favor or compete with superconductivity and how it relates to the spontaneous translation
symmetry breaking such as charge order, still remains controversial.

Since the discovery of the electronic nematicity and its intertwinement with superconductivity in
cuprate superconductors
\cite{Lawler10,Fujita14,Zheng17,Fujita19,Nakata18,Hinkov08,Sato17,Daou10,Taillefer15,Wang21,Ando02,Wu17,Loret19},
the intense efforts have been put forth in order to understand the physical origin of the
electronic nematicity and of its interplay with superconductivity
\cite{Vojta09,Fradkin10,Fernandes19}. As the microscopic origin of the electronic nematicity,
several suggestions have been proposed: the electronic nematicity occurs upon melting of stripe
order or charge order \cite{Kivelson98,Zaanen99,Kivelson03,Nie15}, or arises from the EFS
instability \cite{Halboth00,Yamase06,Kitatani17}, or is attributed to the incommensurate
pair-density-wave \cite{Dai18,Tu19}. In particular, the numerical simulations for some strongly
correlated models indicate that the rotation symmetry-breaking occurs from the underdoped to
overdoped regimes, leading to the EFS distortion \cite{Miyanaga06,Edegger06,Wollny09,Lee16}.
Moreover, these numerical analyses also show that this nematic order competes possibly with
the electron pairing \cite{Miyanaga06,Edegger06,Wollny09}. On the other hand, the interesting theoretical
idea of the nematic-order-driven superconductivity has been put forward, where the fluctuations
associated with the nematic order can enhance superconductivity
\cite{Kitatani17,Maier14,Lederer15,Kaczmarczyk16,Lederer17}. However, up to now, the
intertwinement of the electronic nematicity with superconductivity in cuprate superconductors has
not been discussed starting from a microscopic SC mechanism, and no explicit calculations of the
evolution of $T_{\rm c}$ with the strength of the electronic nematicity has been made so far. In
our recent studies \cite{Feng16,Gao18}, the physical origin of charge order in cuprate
superconductors and of its interplay with superconductivity have been studied based on the
kinetic-energy-driven superconductivity, where the main features of the charge-order state observed
experimentally are qualitatively reproduced. In particular, we \cite{Feng16,Gao18} have shown that
the EFS instability can be interpreted in terms of the electron self-energy effect by which it means
a reconstruction of EFS to form the disconnected Fermi arcs, and then this EFS instability drives
charge-order correlation, with the charge-order wave vector that is well consistent with the wave
vector connecting the tips of the straight Fermi arcs. In this paper, we try to investigate the
intertwinement of the electronic nematicity with superconductivity in cuprate superconductors
along with this line, where we show clearly that in a striking contrast to the role played by
charge order in superconductivity, the emergence of the electronic nematicity favors
superconductivity, with the optimized $T_{\rm c}$ that increases with the increase of the strength
of the electronic nematicity in the weak strength region, and reaches a maximum in the optimal
strength of the electronic nematicity, then decreases with the increase of the strength of the
electronic nematicity in the strong strength region. This enhancement of $T_{\rm c}$ is thus
closely related to the strength of the electronic nematicity, while such an aspect has been
reflected in the electronic structure determined by the single-particle excitations,
where the nematic order breaks the original EFS contour with the four-fold ($C_{4}$) rotation
symmetry to that with a residual two-fold ($C_{2}$) rotation symmetry, and then this EFS contour
with the $C_{2}$ rotation symmetry shrinks down to form the disconnected Fermi arcs with the most
spectral weight that placed at around the tips of the Fermi arcs. In particular, these tips of
the Fermi arcs connected by the scattering wave vectors ${\bf q}_{i}$ construct an {\it octet}
scattering model, however, the partial scattering wave vectors and their respective
symmetry-equivalent partners occur with unequal amplitudes, leading to that these
ordered states driven by the EFS instability break both the rotation and translation symmetries.
As a natural consequence, the electronic structure is inequivalent between the $k_{x}$ and $k_{y}$
directions in momentum space. Moreover, these anisotropic features of the electronic structure
found from the single-particle excitation spectrum are also confirmed via the result of QSI described
in terms of the autocorrelation of the single-particle excitation spectra, where the breaking of
the $C_{4}$ rotation symmetry is verified by the inequivalence on the average of the electronic
structure at the two Bragg scattering sites.

This paper is structured in the following way: The basic formulation is presented in Sec.
\ref{Formalism}, where we generalize the formulation of the kinetic-energy-driven superconductivity
from the previous case without the rotation symmetry-breaking to the present case with broken
rotation symmetry, and then evaluate explicitly the full electron diagonal and off-diagonal propagators
(hence the single-particle excitation spectrum). Based on this framework, we then discuss the phase
diagram of the $T_{\rm c}$ via the strength of the electronic nematicity in Sec. \ref{phase-diagram},
while the exotic features of the electronic structure in the presence of the electronic nematicity
are discussed in Sec. \ref{Renormalization}, where we show that the difference of the spectral
intensities of the single-particle excitation spectrum between the antinodal region near the X and
Y points of the Brillouin zone (BZ) is caused by the electronic nematicity. In Sec.
\ref{collective-response}, we discuss the quantitative characteristics of the rotation symmetry-breaking
of QSI, where the order parameter of the electronic nematicity achieves its maximum in the characteristic
energy, and then decreases rapidly as the energy moves away from the characteristic energy, in
quantitative agreement with the experimental observation \cite{Lawler10,Fujita14,Zheng17,Fujita19}.
Finally, we give a summary and discussions in Sec. \ref{conclusions}. One Appendix is also included.

\section{Model and Formulation}\label{Formalism}

As we have mentioned in Sec. \ref{Introduction}, the basic element of the crystal structure of
cuprate superconductors is the square-lattice CuO$_{2}$ plane \cite{Bednorz86,Wu87}. Immediately
following the discovery of superconductivity in cuprate superconductors, it has been suggested
that the essential physics in the doped CuO$_{2}$ plane can be properly described by the $t$-$J$
model on a square lattice \cite{Anderson87},
\begin{eqnarray}\label{tJ-model}
H&=&-\sum_{<l\hat{\eta}>\sigma}t_{\hat{\eta}}C^{\dagger}_{l\sigma}C_{l+\hat{\eta}\sigma}
+\sum_{<l\hat{\tau}>\sigma}t'_{\hat{\tau}}C^{\dagger}_{l\sigma}C_{l+\hat{\tau}\sigma}
\nonumber\\
&+&\mu\sum_{l\sigma}C^{\dagger}_{l\sigma}C_{l\sigma}
+\sum_{<l\hat{\eta}>}J_{\hat{\eta}}{\bf S}_{l}\cdot {\bf S}_{l+\hat{\eta}},
\end{eqnarray}
where the summations $<l\hat{\eta}>$ and $<l\hat{\tau}>$ are over all sites $l$, and for each site
$l$, restricted to its nearest-neighbor (NN) sites $\hat{\eta}$ and next NN sites $\hat{\tau}$,
respectively, $t_{\hat{\eta}}$ and $t'_{\hat{\tau}}$ denote the electron NN and next NN hoping
amplitudes, and therefore parameterize the kinetic energy, while $J_{\hat{\eta}}$ is the NN magnetic
exchange coupling, and then parameterizes the magnetic energy. This $t$-$J$ model (\ref{tJ-model})
contains both the charge and spin degrees of freedom, and thus describes a competition between the
kinetic energy and magnetic energy. $C^{\dagger}_{l\sigma} (C_{l\sigma}$) is creation (annihilation)
operator for an electron at lattice site $l$ with spin $\sigma$, ${\bf S}_{l}$ is a localized spin
operator at lattice site $l$ with its components $(S^{x}_{l}$, $S^{y}_{l}$, and $S^{z}_{l})$, while
$\mu$ is the electron chemical potential. In cuprate superconductors, the emergence of the
electronic nematicity denotes a state that spontaneously breaks a symmetry of the underlying
$t$-$J$ model (\ref{tJ-model}) on a square lattice which interchanges two axes of the system
\cite{Vojta09,Fradkin10,Fernandes19,Lawler10,Fujita14,Zheng17,Fujita19,Nakata18,Hinkov08,Sato17,Daou10,Taillefer15,Wang21,Ando02,Wu17,Loret19}.
For a better description of the intertwinement of the electronic nematicity with superconductivity,
we introduce an anisotropic parameter $\varsigma$ to represent the orthorhombicity of the NN electron
hoping amplitudes in the $t$-$J$ model (\ref{tJ-model}) as \cite{Nakata18},
\begin{eqnarray}\label{NO-parameter}
t_{\hat{x}}=(1-\varsigma)t, ~~~~ t_{\hat{y}}=(1+\varsigma)t,
\end{eqnarray}
which leads to that the NN anisotropic exchange coupling $J_{\hat{x}}=(1-\varsigma)^{2}J$ and
$J_{\hat{y}}=(1+\varsigma)^{2}J$ in the $t$-$J$ model (\ref{tJ-model}), while the next NN electron
hoping amplitude $t'_{\hat{\tau}}$ is chosen as \cite{Nakata18} $t'_{\hat{\tau}}=t'$. On the one hand,
this band anisotropy introduced via the NN electron hoping amplitudes along the ${\hat{x}}$ and
${\hat{y}}$ directions in Eq. (\ref{NO-parameter}) follows from the previous numerical analyses of the
electronic state in the presence of the electronic nematicity \cite{Edegger06,Wollny09,Lee16}, and
has been also used in the standard tight-binding model to fit the single-particle excitation spectrum
in the presence of the electronic nematicity and the related EFS distortion observed from the ARPES
experiments \cite{Nakata18}. On the other hand, it is possible that the anisotropic parameter
$\varsigma$ in Eq. (\ref{NO-parameter}) is doping and temperature dependence \cite{Nakata18}. In
the present theoretical framework, this anisotropic parameter $\varsigma$ can be also thought to be a
variational parameter, and then in a given doping concentration and a given temperature, this
anisotropic parameter $\varsigma$ can be determined self-consistently by minimizing the energy as the
discussions based on the variational Monte Carlo approach in Ref. \onlinecite{Edegger06}. However, the
experimental observation \cite{Nakata18} and the variational Monte Carlo studies \cite{Edegger06}
have indicated that the order of the magnitude of the variation in the anisotropic parameter
$\varsigma$ is about $0.15\%$ with the error $0.05\%$. As a qualitative discussion in this paper,
this small variation in the anisotropic parameter $\varsigma$ can be neglected, and then the
anisotropic parameter $\varsigma$ is therefore chosen to be independence of doping and temperature in
the following discussions. Moreover, the magnitude of the anisotropic parameter $\varsigma$ also
represents the strength the electronic nematicity in the system. The anisotropic NN electron hoping
amplitudes in Eq. (\ref{NO-parameter}) therefore also indicate that the rotation symmetry is broken
already in the starting $t$-$J$ model (\ref{tJ-model}).

The $t$-$J$ model (\ref{tJ-model}) is defined in a restricted Hilbert space, where the electron
double occupancy is forbidden, i.e., $\sum_{\sigma}C_{l\sigma}^{\dagger}C_{l\sigma}\leq 1$, giving
rise to the strongly electron correlated nature of cuprate superconductors \cite{Anderson87}.
However, the most difficulty for a systematic analysis of the $t$-$J$ model (\ref{tJ-model}) comes
mainly from this no double occupancy constraint \cite{Zhang93,Edegger07}. In order to exclude the
electron double occupancy, we employ the fermion-spin formalism \cite{Feng9404,Feng15}, in which
the electron operators $C_{l\uparrow}$ and $C_{l\downarrow}$ are decoupled as:
$C_{l\uparrow}=h^{\dagger}_{l\uparrow}S^{-}_{l}$ and
$C_{l\downarrow}=h^{\dagger}_{l\downarrow}S^{+}_{l}$, with the spinful fermion operator
$h_{l\sigma}=e^{-i\Phi_{l\sigma}}h_{l}$ that carries the charge of the constrained electron
together with some effects of the spin configuration rearrangements due to the presence of the
doped charge carrier itself, while the spin operator $S_{l}$ carries the spin of the constrained
electron, then the no double occupancy constraint is satisfied in analytical calculations.

Following the $t$-$J$ model in the fermion-spin representation \cite{Feng9404,Feng15}, the
kinetic-energy-driven superconductivity has been established \cite{Feng15,Feng0306,Feng12,Feng15a},
where the interaction between the charge carriers directly from the kinetic energy of the $t$-$J$
model by the exchange of a strongly dispersive {\it spin excitation} is responsible for the d-wave
charge-carrier pairing in the particle-particle channel, then the d-wave electron pairs originated
from the d-wave charge-carrier pairing state are due to the charge-spin recombination, and their
condensation reveals the d-wave SC-state. Based on this kinetic-energy-driven superconductivity,
the interplay between charge order and superconductivity in cuprate superconductors has been
investigated recently \cite{Gao18}, and the results show that although charge order coexists with
superconductivity below $T_{\rm c}$, this charge order competes with superconductivity. Following
up on these previous works on the interplay between charge order and superconductivity, the full
electron diagonal and off-diagonal propagators of the $t$-$J$ model (\ref{tJ-model}) in the
present case with broken rotation symmetry can be evaluated explicitly as [see Appendix
\ref{nematic-order-propagators}],
\begin{subequations}\label{EGF}
\begin{eqnarray}
G_{\varsigma}({\bf k},\omega)&=&{1\over \omega-\varepsilon^{(\varsigma)}_{\bf k}
-\Sigma^{(\varsigma)}_{\rm tot}({\bf k},\omega)}, \label{DEGF}\\
\Im^{\dagger}_{\varsigma}({\bf k},\omega)&=&{L^{(\varsigma)}_{\bf k}(\omega)\over
\omega-\varepsilon^{(\varsigma)}_{\bf k}-\Sigma^{(\varsigma)}_{\rm tot}({\bf k},\omega)},
\label{ODEGF}
\end{eqnarray}
\end{subequations}
where the orthorhombic energy dispersion in the tight-binding approximation is obtained directly
from the $t$-$J$ model (\ref{tJ-model}) as,
\begin{eqnarray}
\varepsilon^{(\varsigma)}_{\bf k}=-4t[(1-\varsigma)\gamma_{{\bf k}_{x}}+(1+\varsigma)
\gamma_{{\bf k}_{y}}]+4t'\gamma_{\bf k}'+\mu,
\end{eqnarray}
with $\gamma_{{\bf k}_{x}}={\rm cos}k_{x}/2$, $\gamma_{{\bf k}_{y}}={\rm cos}k_{y}/2$,
$\gamma_{\bf k}'={\rm cos}k_{x}{\rm cos}k_{y}$, and
\begin{eqnarray}
L^{(\varsigma)}_{\bf k}(\omega)&=&-{\Sigma^{(\varsigma)}_{\rm pp}({\bf k},\omega)\over \omega
+\varepsilon^{(\varsigma)}_{\bf k}+\Sigma^{(\varsigma)}_{\rm ph}({\bf k},-\omega)},
\end{eqnarray}
with $\Sigma^{(\varsigma)}_{\rm ph}({\bf k},\omega)$ and
$\Sigma^{(\varsigma)}_{\rm pp} ({\bf k},\omega)$ that are the normal self-energy in the
particle-hole channel and anomalous self-energy in the particle-particle channel, respectively,
while the total self-energy $\Sigma^{(\varsigma)}_{\rm tot}({\bf k},\omega)$ is a specific
combination of the normal self-energy and anomalous self-energy as,
\begin{eqnarray}
\Sigma^{(\varsigma)}_{\rm tot}({\bf k},\omega)=\Sigma^{(\varsigma)}_{\rm ph}({\bf k},\omega)
+{|\Sigma^{(\varsigma)}_{\rm pp}({\bf k},\omega)|^{2}\over \omega+\varepsilon^{(\varsigma)}_{\bf k}
+\Sigma^{(\varsigma)}_{\rm ph}({\bf k}, -\omega)}. ~~~~~~ \label{TOT-SE}
\end{eqnarray}
In the case of the presence of the electronic nematicity, both the normal self-energy
$\Sigma^{(\varsigma)}_{\rm ph}({\bf k},\omega)$ and anomalous self-energy
$\Sigma^{(\varsigma)}_{\rm pp}({\bf k},\omega)$ still originate from the interaction between
electrons by the exchange of a strongly dispersive spin excitation \cite{Feng15a}, and have been
obtained explicitly in Appendix \ref{nematic-order-propagators}, where the sharp peak visible for
temperature $T\rightarrow 0$ in the normal (anomalous) self-energy is actually a $\delta$-function,
broadened by a small damping used in the numerical calculation at a finite lattice. The calculation
in this paper for the normal (anomalous) self-energy is performed numerically on a $120\times 120$
lattice in momentum space, with the infinitesimal $i0_{+}\rightarrow i\Gamma$ replaced by a small
damping $\Gamma=0.05J$. These full electron diagonal and off-diagonal propagators in Eq.
(\ref{EGF}) and the related total self-energy in Eq. (\ref{TOT-SE}) enable us to systematically
examine the intertwinement of the electronic nematicity with superconductivity.

The electron spectral function $A_{\varsigma}({\bf k},\omega)$ describes the electronic structure
of the system, and can be obtained in terms of the imaginary part of the full electron diagonal
propagator in Eq. (\ref{DEGF}) as,
\begin{eqnarray}\label{ESF}
A_{\varsigma}({\bf k},\omega)={-2{\rm Im}\Sigma^{(\varsigma)}_{\rm tot}({\bf k},\omega)\over
[\omega-\varepsilon^{(\varsigma)}_{\bf k}
-{\rm Re}\Sigma^{(\varsigma)}_{\rm tot}({\bf k},\omega)]^{2}
+[{\rm Im} \Sigma^{(\varsigma)}_{\rm tot}({\bf k},\omega)]^{2}}, ~
\end{eqnarray}
where ${\rm Re}\Sigma^{(\varsigma)}_{\rm tot}({\bf k},\omega)$ and
${\rm Im}\Sigma^{(\varsigma)}_{\rm tot}({\bf k},\omega)$ are the real and imaginary parts of the
total self-energy $\Sigma_{\rm tot}({\bf k},\omega)$, respectively. Since the effect of the
interaction between electrons mediated by a strongly dispersive spin excitation has been encoded
in the total self-energy (then the normal and anomalous self-energies), this leads to two
important modifications of the electron to form the quasiparticle: (i) the dispersion is broadened
by ${\rm Im}\Sigma^{(\varsigma)}_{\rm tot}({\bf k},\omega)$ and (ii) the binding-energy is shifted
by ${\rm Re}\Sigma^{(\varsigma)}_{\rm tot}({\bf k},\omega)$. In this case, the electron spectral
function in Eq. (\ref{ESF}) can be viewed as the probability of the detection of a quasiparticle
with energy $E$ at momentum ${\bf k}$.

The measured photoemission intensity then is directly related to the electron spectral function
$A_{\varsigma}({\bf k},\omega)$, and is given explicitly as
\cite{Damascelli03,Campuzano04,Fink07,Zhou18},
\begin{eqnarray}\label{ARPES-spectrum}
I_{\varsigma}({\bf k},\omega)=|M^{(\varsigma)}_{\rm IF}({\bf k},\omega)|^{2}n_{\rm F}(\omega)
A_{\varsigma}({\bf k},\omega),
\end{eqnarray}
which therefore delivers the information of the electronic structure in a way direct in the
momentum space, where $n_{\rm F}(\omega)$ is the fermion distribution function, while
$M^{(\varsigma)}_{\rm IF}({\bf k},\omega)$ is the dipole matrix element for the single-particle
excitation from the initial to final electronic states in the photoemission process. This dipole
matrix element $M^{(\varsigma)}_{\rm IF}({\bf k},\omega)$ for an initial state of given symmetry
strongly depends on parameters such as the incident photon energy and the light polarization as well
as BZ of the photoemitted electrons. However, this dipole matrix element
$M^{(\varsigma)}_{\rm IF}({\bf k},\omega)$ does not vary significantly with momentum and energy
over the range of the interest \cite{Damascelli03,Campuzano04,Fink07,Zhou18}, and as a qualitative
discussion in this paper, the magnitude of the dipole matrix element
$M^{(\varsigma)}_{\rm IF}({\bf k},\omega)$ can be rescaled to the unit. In the following discussions,
we set the parameters in the $t$-$J$ model (\ref{tJ-model}) as $t/J=3$ and $t'/t=1/3$. However,
when necessary to compare with the experimental data, we choose $J=100$ meV. These are reasonable
model parameters for the phenomenology of cuprate superconductors
\cite{Damascelli03,Campuzano04,Fink07,Zhou18}.

\section{Enhancement of $T_{\rm c}$ by electronic nematicity} \label{phase-diagram}

\begin{figure}[h!]
\centering
\includegraphics[scale=0.95]{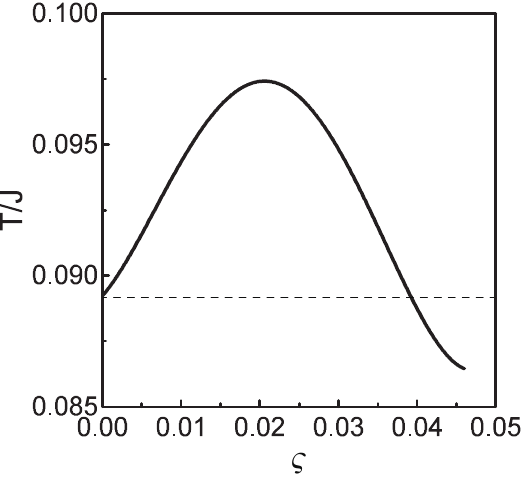}
\caption{The evolution of the superconducting transition temperature with the strength of the
electronic nematicity at the optimal doping $\delta=0.15$ for $t/J=3$ and $t'/t=1/3$.
\label{Tc-nematicity}}
\end{figure}

As we have mentioned in the Sec. \ref{Introduction}, the complexity of the phase diagram of
cuprate superconductors goes well beyond its unique SC-state, as it hosts a variety of different
symmetry-breaking orders \cite{Timusk99,Hufner08,Vishik18,Comin16,Vojta09,Fradkin10,Fernandes19},
such as the nematic order and charge order. From the interplay between charge order and
superconductivity observed experimentally
\cite{Wu11,Tacon12,Comin14,Neto14,Croft14,Hucker14,Campi15,Comin15,Peng16}, it has
been shown that charge order competes with superconductivity \cite{Gao18}. On the other hand,
although the intertwinement of the electronic nematicity with superconductivity has been observed
experimentally
\cite{Vojta09,Fradkin10,Fernandes19,Lawler10,Fujita14,Zheng17,Fujita19,Nakata18,Hinkov08,Sato17,Daou10,Taillefer15,Wang21,Ando02,Wu17},
whether $T_{\rm c}$ can be influenced by the electronic nematicity still is highly debated
\cite{Miyanaga06,Edegger06,Wollny09,Lee16,Maier14,Lederer15,Kaczmarczyk16,Lederer17}. In the case
of the absence of the electronic nematicity, the evolution of $T_{\rm c}$ with the doping
concentration has been obtained in our early studies \cite{Feng15,Feng0306,Feng12,Feng15a} within
the framework of the kinetic-energy-driven superconductivity in terms of the self-consistent
calculation at the condition of the SC gap $\bar{\Delta}=0$, where the maximal $T_{\rm c}$ occurs
around the optimal doping $\delta_{\rm opt}\approx 0.15$, and then decreases in both the underdoped
and overdoped regimes \cite{Feng12,Feng15a}. This dome-like shape doping dependence of $T_{\rm c}$
is in good agreement with the experimental results observed on cuprate superconductors
\cite{Tallon95}. Following these early studies \cite{Feng15,Feng0306,Feng12,Feng15a}, we have
performed a self-consistent calculation for the evolution of $T_{\rm c}$ with the strength of the
electronic nematicity [see Appendix \ref{nematic-order-propagators}], and the result of the
optimized $T_{\rm c}$ as a function of the strength of the electronic nematicity is plotted in
Fig. \ref{Tc-nematicity}. Surprisedly, the exotic change in the optimized $T_{\rm c}$ is closely
associated with the strength of the electronic nematicity from weak to strong, where the
gradual increase in the optimized $T_{\rm c}$ with the increase of the strength of the electronic
nematicity occurs in the weak strength region, and the optimized $T_{\rm c}$ reaches its maximum
in the optimal strength of the electronic nematicity $\varsigma\approx 0.022$, subsequently, the
optimized $T_{\rm c}$ decreases gradually with the increase of the strength of the electronic
nematicity in the strong strength region, in a striking similar to the dome-like shape doping
dependence of $T_{\rm c}$. This dome-like shape nematic-order strength dependence of $T_{\rm c}$
therefore shows clearly that superconductivity in cuprate superconductors is enhanced by the
electronic nematicity. Experimentally, from the anisotropy of the magnetic susceptibility of
YBa$_{2}$Cu$_{3}$O$_{7-y}$ observed by the magnetic torque measurements \cite{Sato17,Daou10}, the
range of the strength of the electronic nematicity was estimated as $\varsigma>0.005$, while the
order of the magnitude of the strength $\varsigma$ with the high impacts on various physical
properties has been estimated as $\varsigma\sim 0.01$ in the ARPES measurements \cite{Nakata18}
of the anisotropy of the electronic structure of
Bi$_{1.7}$Pb$_{0.5}$Sr$_{1.9}$CaCu$_{2}$O$_{8+\delta}$. It is thus shown that the actual range of
the strength of the electronic nematicity that $T_{\rm c}$ is enhanced in Fig. \ref{Tc-nematicity}
is well similar in theory and experiments. However, it should be also noted that in the extremely
strong strength region $\varsigma> 0.04$, the optimized $T_{\rm c}$ is lower than that in the case
of the absence of the electronic nematicity, although the experimental data of the reduction of
superconductivity in the extremely strong strength region is still lacking to date.

The enhancement of superconductivity by the electronic nematicity in cuprate superconductors
implies that the energy in the SC-state with coexisting electronic nematicity is lower than the
corresponding energy in the SC-state with the absence of the electronic nematicity, i.e., the
SC-state with coexisting electronic nematicity is more stable than the SC-state with the absence
of the electronic nematicity. In Appendix \ref{nematic-order-propagators}.3, the evolution of the
charge carrier pair gap parameter (then the electron pair gap parameter in
Appendix \ref{nematic-order-propagators}.6) with the strength of the electronic nematicity has
been evaluated self-consistently, and the result of the maximal charge-carrier pair gap parameter
$\bar{\Delta}^{\rm (h)}_{\varsigma{\rm max}}$ as a function of the strength of the electronic
nematicity is plotted in Fig. \ref{pair-gap-parameter-nematicity}, where with the increase of the
strength of the electronic nematicity, $\bar{\Delta}^{\rm (h)}_{\varsigma{\rm max}}$ increases in
the weak strength region, and achieves its maximum at around the same {\it optimal} strength of
the electronic nematicity $\varsigma\approx 0.022$ as the nematic-order strength dependence of
$T_{\rm c}$ shown in Fig. \ref{Tc-nematicity}. However, with the further increase of the strength,
$\bar{\Delta}^{\rm (h)}_{\varsigma{\rm max}}$ turns into a decrease in the strong strength region.
Moreover, we \cite{Cao22,Zhao12} have also evaluated the ground-state energies both in the SC-state
and normal-state with coexisting electronic nematicity, and found that the condensation energy
$\Delta E^{(\varsigma)}_{\rm g}$ (then the energy difference between the normal-state energy and
SC-state energy) is proportional to the charge-carrier pair gap parameter, i.e.,
$\Delta E^{(\varsigma)}_{\rm g}=E^{\rm (n)}_{\varsigma{\rm g}}-E^{\rm (s)}_{\varsigma{\rm g}}\propto
\bar{\Delta}^{\rm (h)}_{\varsigma{\rm max}}$, which therefore shows that (i) the energy in the
SC-state with coexisting electronic nematicity is lower than the SC-state energy with the absence of
the electronic nematicity, indicating that the electronic nematicity together with the associated
fluctuation phenomena in cuprate superconductors are a natural consequence of the strong electron
correlation effect; (ii) the lowest energy in the SC-state with coexisting electronic nematicity
occurs at around the same {\it optimal} strength of the electronic nematicity
$\varsigma\approx 0.022$. This nematic-order strength dependence of the condensation energy and the
related nematic-order strength dependence of the pair gap parameter thus explain why
superconductivity is enhanced by the electronic nematicity, and $T_{\rm c}$ exhibits a dome-like shape
nematic-order strength dependence. Furthermore, $T_{\rm c}$ as a function of the strength of the
electronic nematicity at the doping $\delta=0.06$ has been obtained very recently in Ref. \onlinecite{Cao21}.
Incorporating the present result in Fig. \ref{Tc-nematicity} with that obtained in Ref. \onlinecite{Cao21},
it is thus shown that although $T_{\rm c}$ for a given strength of the electronic nematicity at the
doping $\delta=0.06$ is much lower than the corresponding $T_{\rm c}$ at the optimal doping
$\delta=0.15$ shown in Fig. \ref{Tc-nematicity}, the global dome-like shape of the nematic-order
strength dependence of the enhancement of $T_{\rm c}$ at the doping $\delta=0.06$ together with the
magnitude of the {\it optimal} strength obtained in Ref. \onlinecite{Cao21} is the same with that at the
optimal doping $\delta=0.15$ shown in Fig. \ref{Tc-nematicity}, which therefore show that the
enhancement of superconductivity by the electronic nematicity occurs at an any given doping
concentration of the SC dome.

\section{Electronic structure in superconducting-state with coexisting electronic nematicity}
\label{Renormalization}

The electronic nematicity emerged as a key feature of cuprate superconductors has high impacts on
various properties
\cite{Vojta09,Fradkin10,Fernandes19,Lawler10,Fujita14,Zheng17,Fujita19,Nakata18,Hinkov08,Sato17,Daou10,Taillefer15,Wang21,Ando02,Wu17,Loret19}.
In this section, we analyze the electronic structure of cuprate superconductors in the SC-state
with coexisting electronic nematicity to shed light on the exotic features of the quasiparticle
excitations. This nematic order coexist with spontaneous translation symmetry-breaking such as
charge order leading to that the multiple modulations for the SC-state electronic structure occur
simultaneously.

\subsection{Rotation symmetry-breaking of electron Fermi surface}\label{two-fold-EFS}

Angle-resolved photoemission spectroscopy (ARPES) experiments
\cite{Damascelli03,Campuzano04,Fink07,Zhou18} measure the single-particle excitation spectrum,
while the underlying EFS contour in momentum space is directly obtained by the trace of the peak
positions in the single-particle excitation spectrum. In the absence of any sort of
symmetry-breaking, the shape of EFS of cuprate superconductors reflects the underlying symmetry of
the square-lattice CuO$_{2}$ plane. In particular, the shape of EFS has deep consequences for the
low-energy properties \cite{Timusk99,Hufner08,Vishik18,Comin16,Vojta09,Fradkin10,Fernandes19}, and
has been also central to addressing multiple electronic orders
\cite{Vojta09,Fradkin10,Fernandes19,Lawler10,Fujita14,Zheng17,Fujita19,Nakata18,Hinkov08,Sato17,Daou10,Taillefer15,Wang21,Ando02,Wu17,Loret19}.
This is why the determination of the shape of EFS in cuprate superconductors is believed to be key
issue for the understanding of the physical origin of different electronic ordered
(then density-wave) states and of their intimate interplay with superconductivity.

In the absence of the rotation symmetry-breaking, we \cite{Gao18,Feng15a} have shown within the
framework of the kinetic-energy-driven superconductivity that although the EFS contour exhibits
a $C_{4}$ rotation symmetry on the square lattice, the momentum dependence of the quasiparticle
scattering due to the electron interaction by the exchange of a strongly dispersive spin
excitation breaks up the EFS contour into the disconnected Fermi arcs with the most spectral
weight that accommodates at around the tips of the Fermi arcs to form the Fermi-arc-tip liquid.
As a natural consequence of the $C_{4}$ rotation symmetry on the square lattice, the two tips of
the Fermi arc in each quarter of BZ are symmetrical about the nodal (diagonal) direction.
Moreover, these tips of the Fermi arcs connected by the scattering wave vectors ${\bf q}_{i}$
construct an {\it octet} scattering model, where for any quasiparticle scattering process, the
scattering wave vector and its symmetry-equivalent partner occur with equal amplitude. In this
case, a bewildering variety of electronically ordered states with the translation
symmetry-breaking described by the quasiparticle scattering processes with the corresponding
scattering wave vectors ${\bf q}_{i}$ then are driven by this EFS instability
\cite{Vishik18,Comin16,Vojta09,Wu11,Tacon12,Comin14,Neto14,Croft14,Hucker14,Campi15,Comin15,Peng16}.
Although these electronically ordered states do not break the $C_{4}$ rotation symmetry on the
square lattice, they break the discrete translation symmetry
\cite{Vishik18,Comin16,Vojta09,Wu11,Tacon12,Comin14,Neto14,Croft14,Hucker14,Campi15,Comin15,Peng16}.

\begin{figure}[h!]
\centering
\includegraphics[scale=0.95]{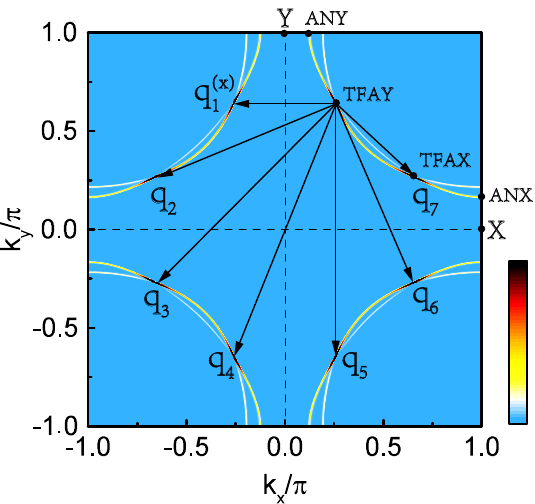}
\caption{(Color online) The electron Fermi surface map at $\delta=0.15$ with $T=0.002J$ in the optimal
strength of the electronic nematicity $\varsigma=0.022$ for $t/J=3$ and $t'/t=1/3$, where ANX (ANY) and
TFAX (TFAY) denote the antinode and tip of the Fermi arc near the X (Y) point of the Brillouin zone,
respectively, and ${\bf q}_{1}$, ${\bf q}_{2}$, ${\bf q}_{3}$, ${\bf q}_{4}$, ${\bf q}_{5}$,
${\bf q}_{6}$, and ${\bf q}_{7}$ indicate different scattering wave vectors. However, the scattering
wave vectors ${\bf q}_{1}$, ${\bf q}_{4}$, and ${\bf q}_{5}$ and the respective symmetry-corresponding
partners occur with unequal amplitudes. \label{EFS-map}}
\end{figure}

However, the original $C_{4}$ rotation symmetry in EFS is broken in the presence of the electronic
nematicity. To understand this EFS anisotropy more clearly, we plot the map of the single-particle
excitation spectral intensity $I({\bf k},0)$ in Eq. (\ref{ARPES-spectrum}) at doping $\delta=0.15$
with temperature $T=0.002J$ in the optimal strength of the electronic nematicity $\varsigma=0.022$
in Fig. \ref{EFS-map}. Apparently, the electronic nematicity corresponds down to the $C_{2}$
rotation symmetry, where the most exotic features can be summarized as: (i) the original EFS with
the $C_{4}$ rotation symmetry on the square lattice in the absence of the electronic nematicity
is broken up into that with a residual $C_{2}$ rotation symmetry, which leads to the EFS distortion.
As a natural consequence, the electronic structure at around the antinode near the X point (ANX) of
BZ is inequivalent to that at around the antinode near the Y point (ANY). In particular, with the
increase of the strength of the electronic nematicity, the antinode near the X point of BZ shifts
away from the $[\pi,0]$ point, while the antinode near the Y point moves towards to the $[0,\pi]$
point, in qualitative agreement with the ARPES experimental results \cite{Nakata18}. Moreover, as in
the case of the absence of the electronic nematicity \cite{Gao18}, a bifurcation for the original
single-contour EFS occurs due to the renormalization of the electrons in the presence of the strong
electron interaction mediated by a strongly dispersive spin excitation except for at around the tips
of the Fermi arcs, where this bifurcation disappears. However, the spectral weight on the extra sheet
is suppressed, leading to that this extra sheet become unobservable in experiments; (ii) although the
special feature of the most spectral weight that  accommodates at around the tips of the Fermi arcs
remains, the symmetrical relation of two tips of the Fermi arc in each quarter of BZ about the nodal
(diagonal) direction vanishes; (iii) concomitantly, the original $C_{4}$ rotation symmetry for the
{\it octet} scattering model in the absence of the electronic nematicity is broken into the $C_{2}$
rotation symmetry, where for the partial quasiparticle scattering processes with the corresponding
scattering wave vectors ${\bf q}_{1}$, ${\bf q}_{4}$, and ${\bf q}_{5}$, the amplitudes of the
scattering wave vectors are respectively inequivalent to their symmetry-corresponding partners.
This EFS instability in the presence of the electronic nematicity then drives multiple ordered
states, where the electronic ordered states described by the quasiparticle scattering processes with
the corresponding scattering wave vectors ${\bf q}_{1}$, ${\bf q}_{4}$, and ${\bf q}_{5}$ break both
the rotation and translation symmetries, while these ordered states described by the quasiparticle
scattering processes with the corresponding scattering wave vectors ${\bf q}_{2}$, ${\bf q}_{3}$,
${\bf q}_{6}$, and ${\bf q}_{7}$ only break the translation symmetry. In particular, the charge-order
wave vector $Q^{(x)}_{\rm co}={\bf q}^{(x)}_{1}$ is on amplitude an inequality with its
symmetry-corresponding partner $Q^{(y)}_{\rm co}={\bf q}^{(y)}_{1}$, i.e.,
$Q^{(x)}_{\rm co}\neq Q^{(y)}_{\rm co}$, leading to the emergence of the most significant stripe type
charge order with broken both rotation and translation symmetries \cite{Comin15a,Zhang18}, i.e., this
stripe type charge order has a nematic character. Furthermore, as in the case without the rotation
symmetry-breaking \cite{Gao18}, the positions of the tips of the Fermi arcs are doping dependent,
which therefore leads to the evolution of the amplitudes of the scattering wave vectors ${\bf q}_{i}$
with doping. More specifically, the magnitude of the charge-order wave vector $Q_{\rm co}$ smoothly
decreases with the increase of doping \cite{Gao18}, in qualitative agreement with the experimental data
\cite{Vishik18,Comin16,Vojta09,Wu11,Tacon12,Comin14,Neto14,Croft14,Hucker14,Campi15,Comin15,Peng16}.
This {\it octet} scattering model with the $C_{2}$ rotation symmetry shown in
Fig. \ref{EFS-map} is a fundamental quasiparticle scattering model \cite{Wang03,Gao19} in the
explanation of the rotation symmetry-breaking of the QSI experimental data
\cite{Lawler10,Fujita14,Zheng17,Fujita19}, and also can give a consistent description of the regions
of the highest joint density of states observed from the ARPES autocorrelation experiments
\cite{Wang03,Gao19,Chatterjee06,He14}. We will return to this discussion towards
Sec. \ref{collective-response} of this paper.

Within the framework of the kinetic-energy-driven superconductivity
\cite{Feng15,Feng0306,Feng12,Feng15a}, the SC quasiparticles are the phase-coherent linear
superpositions of electrons and holes as in the conventional superconductors, while the SC condensate
is made up of electron pairs, which are bound-states of two electrons with opposite momenta and spins.
On the other hand, the electronically ordering quasiparticles are superpositions of electrons (or holes),
and then the electronically ordered state is likewise a pair condensate, but of electrons and holes, whose
net momentum determines the wave-length of electronic ordering \cite{Gao18,Kohn70,Hinton16}. In the
presence of the electronic nematicity, the pairing of electrons and holes at ${\bf k}$ and
${\bf k}+{\bf q}_{i}$ separated by the electronic order wave vector ${\bf q}_{i}$ drives the
symmetry-breaking ordered state formation, whereas the electron pairing at ${\bf k}$ and ${-\bf k}$
states induces superconductivity. This is why the SC-state of cuprate superconductors coexists with a
bewildering variety of the electronically ordered states with the symmetry-breaking, leading to that the
multiple modulations for the SC-state occur simultaneously.

\begin{figure}[h!]
\centering
\includegraphics[scale=0.70]{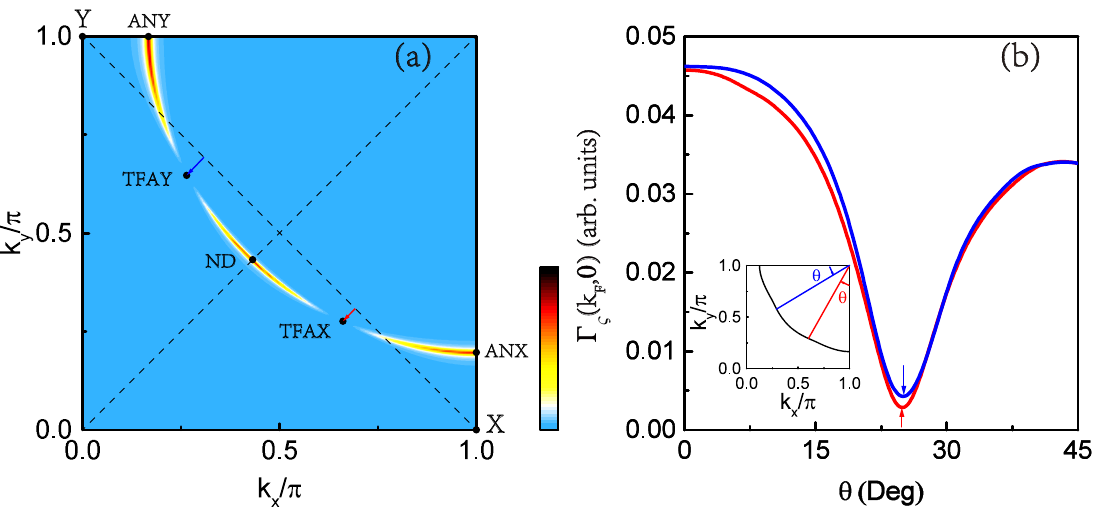}
\caption{(Color online) (a) The intensity map of the quasiparticle scattering rate and (b) the angular
dependence of the quasiparticle scattering rate along the electron Fermi surface from the antinode near
the X point of the Brillouin zone to the node (red line) and from the antinode near the Y point to the
node (blue line) at $\delta=0.15$ with $T=0.002J$ in the optimal strength of the electronic nematicity
$\varsigma=0.022$ for $t/J=3$ and $t'/t=1/3$, where ANX (ANY) and TFAX (TFAY) in (a) denote the positions
of the antinode and tip of the Fermi arc near the X (Y) point, respectively, and ND denotes the position
of the node, while the red (blue) arrow in (b) indicates the corresponding minimum of the quasiparticle
scattering rate (then the corresponding position of the tip of the Fermi arc). \label{scattering-rate}}
\end{figure}

The above obtained $C_{4}$ rotation symmetry-breaking of EFS is caused by the electronic nematicity,
while the EFS reconstruction to form the disconnected Fermi arcs is still attributed to the momentum
dependence of the quasiparticle scattering rate \cite{Feng16,Gao18}
$\Gamma_{\varsigma}({\bf k},0)={\rm Im}\Sigma^{(\varsigma)}_{\rm tot}({\bf k},0)$. This follows a basic
fact that in despite of the presence of the electronic nematicity, the EFS contour in momentum space is
determined directly by the poles of the full electron diagonal propagator (\ref{DEGF}) at zero energy,
\begin{eqnarray}\label{EFS-contour}
\varepsilon^{(\varsigma)}_{{\bf k}_{\rm F}}-{\rm Re}\Sigma^{(\varsigma)}_{\rm tot}({\bf k}_{\rm F},0)=0,
\end{eqnarray}
and then the spectral weight on the EFS contour is dominated mainly by the inverse of the quasiparticle
scattering rate $1/\Gamma_{\varsigma}({\bf k}_{\rm F},0)$. However, the electronic nematicity generates
a breaking of the symmetrical relation of $\Gamma_{\varsigma}({\bf k}_{\rm F},0)$ about the nodal
(diagonal) direction. To pinpoint this origin of EFS with broken rotation symmetry, we plot (a) the
intensity map of $\Gamma_{\varsigma}({\bf k}_{\rm F},0)$ and (b) the angular dependence of
$\Gamma_{\varsigma}({\bf k}_{\rm F},0)$ at $\delta=0.15$ with $T=0.002J$ in the optimal strength of the
electronic nematicity $\varsigma=0.022$ in Fig. \ref{scattering-rate}, where the three exotic features
emerge as: (i) The quasiparticle scattering is stronger at around the antinodal region than at around
the nodal region. In this case, the spectral weight on the EFS contour ${\bf k}_{\rm F}$ at around the
antinodal region is substantially suppressed, while the spectral weight at around the nodal region is
reduced modestly, which leads to form the disconnected Fermi arcs around the nodal region; (ii) The
weakest quasiparticle scattering does not occur at around the node, but occurs exactly at the tips of the
Fermi arcs, which further reduces the most part of the spectral weight on the Fermi arcs to the tips of
the Fermi arcs. This angular dependences of $\Gamma_{\varsigma}({\bf k}_{\rm F},0)$ is in qualitative
agreement with the experimental data \cite{Vishik09}, where the weakest quasiparticle scattering
appeared at around the tips of the Fermi arcs has been observed. These tips of the Fermi arcs connected
by the scattering wave vectors ${\bf q}_{i}$
therefore construct an {\it octet} scattering model; (iii) However, the positions of the strongest
scattering at the antinode and the weakest scattering at the tip of the Fermi arc near X point of BZ are
clearly different from the corresponding positions of the strongest scattering at the antinode and the
weakest scattering at the tip of the Fermi arc near Y point, and then the symmetrical relation of
$\Gamma_{\varsigma}({\bf k}_{\rm F},0)$ about the nodal (diagonal) direction disappears. This anisotropy
of $\Gamma_{\varsigma}({\bf k}_{\rm F},0)$ therefore induces EFS with broken $C_{4}$ rotation symmetry
shown in Fig. \ref{EFS-map}. Moreover, the positions of the weakest quasiparticle scattering is doping
dependent, which therefore leads to the positions of the tips of the Fermi arcs (then the scattering
wave vectors ${\bf q}_{i}$ in the octet scattering model) are the evolution with doping.

\subsection{Line-shape anisotropy of energy distribution curve}

We now turn to our attention to the anisotropy of the electronic structure of cuprate superconductors
in the SC-state with coexisting electronic nematicity. The most intuitive approach is to analyze section
corresponding to a fixed momentum ${\bf k}$, i.e., the energy distribution curve, where the
characteristic feature is so-called peak-dip-hump (PDH) structure
\cite{Dessau91,Randeria95,Norman97,Campuzano99,DLFeng02,Wei08,Hashimoto15,DMou17}. This remarkable PDH
structure consists of a sharp quasiparticle excitation peak at the lowest binding-energy, a broad hump at
the higher binding-energy, and a spectral {\it dip} between them. Theoretical, there is a general
consensus that the emergence of the {\it dip} is a natural consequence of very strong scattering of the
electrons mediated by {\it bosonic excitations}, although what type bosonic excitation that is the
appropriate bosonic excitation for the role of the electron pairing glue is still under debate. In
particular, both the experimental and theoretical studies indicate that the sharp peak in the
quasiparticle scattering rate is directly responsible for the outstanding PDH structure in the energy
distribution curve \cite{DMou17,Gao18a,Liu20a}.

\begin{figure}[h!]
\centering
\includegraphics[scale=0.85]{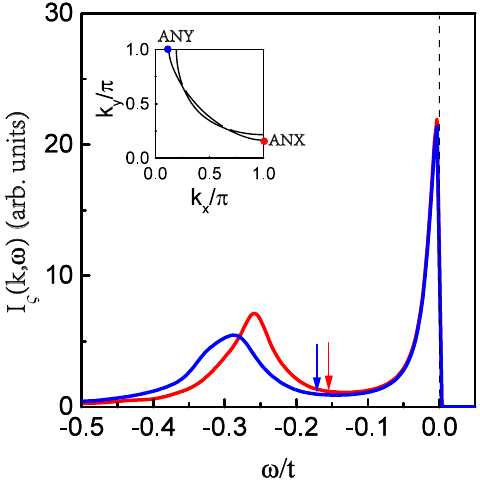}
\caption{(Color online) The energy distribution curves at the antinode near the X point of the Brillouin
zone (red line) and at the antinode near the Y point (blue line) in $\delta=0.15$ with $T=0.002J$ and the
optimal strength of the electronic nematicity $\varsigma=0.022$ for $t/J=3$ and $t'/t=1/3$, where the
arrows indicate the positions of the dips, while ANX and ANY in the inset denote the antinodes near the X
and Y points, respectively.  \label{PDH-structure}}
\end{figure}

In the absence of the electronic nematicity, the line-shape of the energy distribution curve exhibits the
$C_{4}$ rotation symmetry on the square lattice, reflecting a basic fact that the electronic structure is
equivalent between the $k_{x}$ and $k_{y}$ directions in momentum space
\cite{Dessau91,Randeria95,Norman97,Campuzano99,DLFeng02,Wei08,Hashimoto15,DMou17}. However, in
corresponding to the $C_{4}$ rotation symmetry-breaking of EFS shown in Fig. \ref{EFS-map}, the electronic
nematicity induces the line-shape anisotropy of the energy distribution curve. To see this point more
clearly, we plot the energy distribution curves at the antinode near the X point of BZ (red line) and at
the antinode near the Y point (blue line) for $\delta=0.15$ with $T=0.002J$ in the optimal strength of the
electronic nematicity $\varsigma=0.022$ in Fig. \ref{PDH-structure}, where although the global feature of
the PDH structure in the energy distribution curve is in a striking analogy to that in the corresponding
case without the rotation symmetry-breaking, the line-shape of the energy distribution curve at the
antinode near the X point of BZ is not identical with that at the antinode near the Y point, leading to
the line-shape anisotropy of the energy distribution curve. In particular, the spectral intensities are
higher at the antinode near X point than at the antinode near Y point, in agreement with the ARPES
experimental results \cite{Nakata18}. Moreover, the hump and dip energies in the PDH structure at the
antinode near the X point is lower than the corresponding hump and dip energies at the antinode near the
Y point, respectively, which also conform to the result of EFS with the $C_{2}$ rotation symmetry shown
in Fig. \ref{EFS-map}. However, it should be noted that although the electronic structure becomes
inequivalent between the $k_{x}$ and $k_{y}$ directions, the electronic structure along the nodal direction
is unaffected, also in agreement with the ARPES experimental results \cite{Nakata18}.

\begin{figure}[h!]
\centering
\includegraphics[scale=0.85]{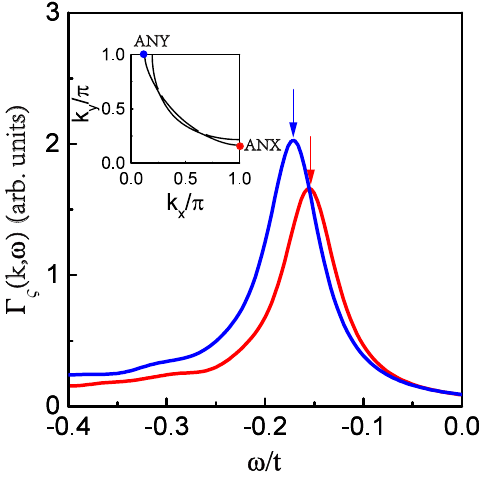}
\caption{(Color online) The quasiparticle scattering rate as a function of energy at the
antinode near the X point of the Brillouin zone (red line) and at the antinode near the Y
point (blue line) in $\delta=0.15$ with $T=0.002J$ and the optimal strength of the
electronic nematicity $\varsigma=0.022$ for $t/J=3$ and $t'/t=1/3$, where the arrows
indicate the positions of the peaks, while ANX and ANY in the insets denote the antinodes
near the X and Y points, respectively. \label{scattering-rate-PDH}}
\end{figure}

This inequivalence of the electronic structure between the $k_{x}$ and $k_{y}$ directions associated with
the electronic nematicity can be also interpreted in terms of the quasiparticle scattering rate anisotropy.
The origin of the emergence of the PDH structure in the energy distribution curve is intimately related to
the corresponding peak-structure in the quasiparticle scattering rate generated from the electron
interaction by the exchange of a strongly dispersive spin excitation \cite{DMou17,Gao18a,Liu20a}. From the
single-particle excitation spectrum in Eq. (\ref{ARPES-spectrum}), the position of the peak in the energy
distribution curve is determined by the renormalized quasiparticle excitation energy in terms of the
self-consistent equation,
\begin{eqnarray}\label{excitation-energy}
\omega-\varepsilon^{(\varsigma)}_{\bf k}-{\rm Re}\Sigma^{(\varsigma)}_{\rm tot}({\bf k},\omega)=0,
\end{eqnarray}
and then the intensity of this peak is dominated by the inverse of the quasiparticle scattering rate
$\Gamma_{\varsigma}({\bf k},\omega)$ [then the imaginary part of the total self-energy
${\rm Im}\Sigma^{(\varsigma)}_{\rm tot}({\bf k},\omega)$]. In other words, the spectral line-shape in the
energy distribution curve is determined by both the real and imaginary parts of the total self-energy.
However, in the presence of the electronic nematicity, the peak energy of
$\Gamma_{\varsigma}({\bf k},\omega)$ is lower at the antinode near the X point than at the antinode near
the Y point, leading to the inequivalence of the line-shape between the antinodes near X and Y points.
To test this idea directly, we plot the quasiparticle scattering rate $\Gamma_{\varsigma}({\bf k},\omega)$
as a function of energy at the antinode near the X point (red line) and at the antinode near the Y point
(blue line) for $\delta=0.15$ with $T=0.002J$ in the optimal strength of the electronic nematicity
$\varsigma=0.022$ in Fig. \ref{scattering-rate-PDH}, where $\Gamma_{\varsigma}({\bf k},\omega)$ exhibits
a sharp peak-structure at both the antinodes near the X and Y points, with the position of the peak in
$\Gamma_{\varsigma}({\bf k},\omega)$ at the antinode near the X (Y) point corresponds exactly to the
position of the dip in the PDH structure at the antinode near the X (Y) point shown in
Fig. \ref{PDH-structure}, indicating that the peak-structure in $\Gamma_{\varsigma}({\bf k},\omega)$ is
responsible directly for the PDH structure in the energy distribution curve \cite{DMou17,Gao18a,Liu20a}.
However, the peak energy in $\Gamma_{\varsigma}({\bf k},\omega)$ at the antinode near the X point is
different from that at the antinode near the Y point, which leads to the inequivalence of the line-shape
in the energy distribution curve between the antinodes near X and Y points. Moreover, the intensity of
the peak in $\Gamma_{\varsigma}({\bf k},\omega)$ at the antinode near the Y point is higher than that at
the antinode near the X point, this is why the spectral intensities shown in Fig. \ref{PDH-structure}
are higher at the antinode near X point than at the antinode near Y point.

\section{Quantitative characteristics of rotation symmetry-breaking of quasiparticle scattering
interference} \label{collective-response}

The exploration of QSI in cuprate superconductors can be used to elucidate the nature of the quasiparticle
excitation and of its interplay with symmetry-breaking orders and superconductivity. This follows an
experimental fact that the quasiparticles scattering from impurities interfere with one another, producing
a notable QSI pattern in the inhomogeneous part for the Fourier transform (FT) of the scanning tunneling
spectroscopy (STS) measured real-space image of the local density of states (LDOS)
\cite{Yin21,Pan01,Hoffman02,Kohsaka07,Kohsaka08,Hamidian16}. From these observed QSI patterns as a function
of energy, one can obtain the energy variation of the momentum ${\bf q}$, which is an autocorrelation
between the electronic bands $E_{\bf k}$ and $E_{{\bf k}+{\bf q}}$. For the same band $E_{\bf k}$, the
intensity distribution of the QSI pattern can be different, depending on the quasiparticle scattering
geometry. However, this quasiparticle scattering geometry is closely associated with the shape of EFS and
the related electron distribution, which are directly related to the single-particle excitation
spectrum. In other words, the QSI intensity is proportional to the intensities of the single-particle
excitation spectra at the momenta ${\bf k}$ and ${\bf k}+{\bf q}$, while the intensity peaks in the QSI
pattern then are corresponding to the highest joint density of states. This is why the dispersion of the
peaks in the QSI pattern as a function of energy is analyzed in terms of the octet scattering model shown
in Fig. \ref{EFS-map} and yields the crucial information about the shape of EFS and the related nature of
the quasiparticle excitation. Moreover, by the measurement of the characteristic feature of the Bragg peaks
in FT of a real-space image of LDOS, one is considering the phenomena that occur with the periodicity of
the underlying square lattice and which qualify any rotation symmetry-breaking. On the other hand, the
ARPES autocorrelation \cite{Chatterjee06,He14},
\begin{eqnarray}\label{ARPES-autocorrelation}
{\bar C}_{\varsigma}({\bf q},\omega)={1\over N}\sum_{\bf k}I_{\varsigma}({\bf k}+{\bf q},\omega)
I_{\varsigma}({\bf k},\omega),
\end{eqnarray}
measures the autocorrelation of the single-particle excitation spectra in Eq. (\ref{ARPES-spectrum}) at
two different momenta ${\bf k}$ and ${\bf k}+{\bf q}$, where the summation of momentum ${\bf k}$ is
extended up to the second BZ for the discussion of QSI together with the Bragg scattering
\cite{Chatterjee06}, and $N$ is the number of lattice sites. This ARPES autocorrelation in
Eq. (\ref{ARPES-autocorrelation}) therefore detects the effectively momentum-resolved joint density of
states in the electronic state, and can give us the important insights into the nature of the
quasiparticle excitation. In particular, the experimental observations \cite{Chatterjee06,He14} have
demonstrated that the ARPES autocorrelation exhibits discrete spots in momentum-space, which are directly
related with the quasiparticle scattering wave vectors ${\bf q}_{i}$ connecting the tips of the Fermi
arcs in the octet scattering model shown in Fig. \ref{EFS-map}, and are also well consistent with the
QSI peaks observed from the FT-STS experiments \cite{Yin21,Pan01,Hoffman02,Kohsaka07,Kohsaka08,Hamidian16}.
Moreover, we \cite{Gao19} have also shown within the framework of the kinetic-energy-driven
superconductivity that there is an intimate connection of the ARPES autocorrelation with QSI in cuprate
superconductors, where the inhomogeneous part for FT LDOS in the presence of a single point-like impurity
scattering potential has been evaluated based on the octet scattering model, and then the obtained
momentum-space structure of the QSI patterns in the presence of the strong scattering potential is well
consistent with the momentum-space structure of the ARPES autocorrelation patterns. In other words, the
{\it octet} scattering model with the quasiparticle scattering wave vectors ${\bf q}_{i}$ connecting the
corresponding tips of the Fermi arcs that can give a consistent description of the regions of the highest
joint density of states in the ARPES autocorrelation can be also used to explain the QSI data observed
from the FT-STS experiments. This is also why the main features of QSI
\cite{Yin21,Pan01,Hoffman02,Kohsaka07,Kohsaka08,Hamidian16} can be also obtained in terms of the ARPES
autocorrelation \cite{Chatterjee06,He14}. However, the original electronic structure with the $C_{4}$
rotation symmetry is thus broken in the presence of the electronic nematicity, while such an aspect should
be reflected in the corresponding QSI. In this subsection, we discuss the characteristic feature of the
rotation symmetry-breaking of QSI in terms of the ARPES autocorrelation, and then make a comparison with
the results of the rotation symmetry-breaking QSI measured from the FT-STS experiments
\cite{Lawler10,Fujita14,Zheng17,Fujita19}.

\begin{figure}[h!]
\centering
\includegraphics[scale=0.85]{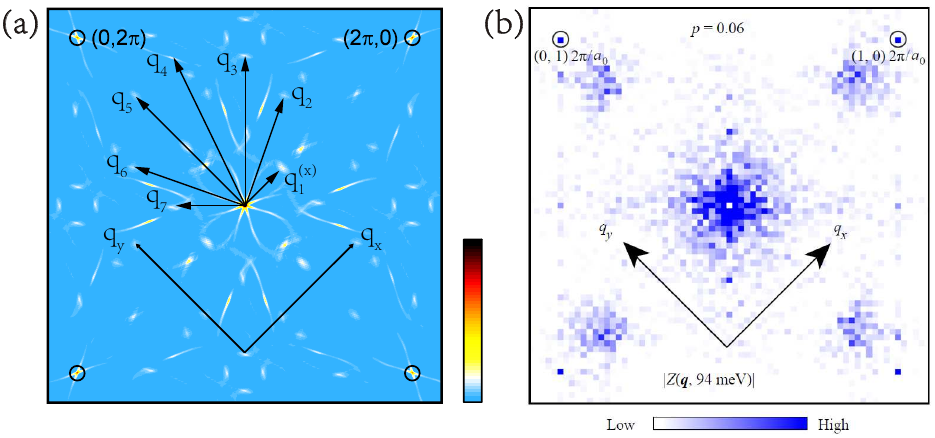}
\caption{(Color online) (a) The intensity map of the ARPES autocorrelation in a $[q_{x},q_{y}]$ plane at the
binding-energy $\omega=94$ meV and $\delta=0.06$ with $T=0.002J$ and the optimal strength of the electronic
nematicity $\varsigma=0.022$ for $t/J=3$ and $t'/t=1/3$. ${\bf q}_{1}$, ${\bf q}_{2}$, ${\bf q}_{3}$,
${\bf q}_{4}$, ${\bf q}_{5}$, ${\bf q}_{6}$, and ${\bf q}_{7}$ in (a) indicate different scattering wave
vectors, however, the scattering wave vectors ${\bf q}_{1}$, ${\bf q}_{4}$, and ${\bf q}_{5}$ and the
respective symmetry-corresponding partners occur with unequal amplitudes. The locations of the Bragg peaks
${\bf Q}^{\rm (B)}_{x}$ and ${\bf Q}^{\rm (B)}_{y}$ in (a) are indicated by circles. (b) The corresponding
experimental result of the quasiparticle scattering interference pattern for
Bi$_{2}$Sr$_{2}$CaCu$_{2}$O$_{8+\delta}$ at doping $\delta=0.06$ in the binding-energy $\omega=94$ meV taken
from Ref. \onlinecite{Fujita19}. \label{autocorrelation-maps}}
\end{figure}

\begin{figure}[h!]
\centering
\includegraphics[scale=0.75]{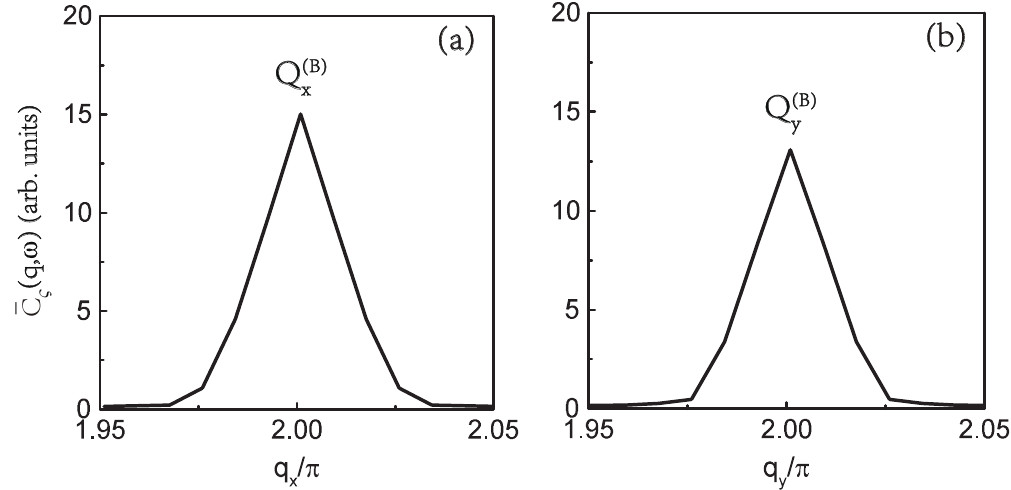}
\caption{The intensity of the peak as a function of momentum (a) at around the Bragg wave vector
${\bf Q}^{\rm (B)}_{x}$ along the $q_{x}$ direction and (b) at around the Bragg wave vector
${\bf Q}^{\rm (B)}_{y}$ along the $q_{y}$ direction in the binding-energy $\omega=94$ meV and
$\delta=0.06$ with $T=0.002J$ and the optimal strength of the electronic nematicity $\varsigma=0.022$
for $t/J=3$ and $t'/t=1/3$.  \label{Bragg-peak}}
\end{figure}

Now we are ready to discuss the characteristic feature of QSI together with the Bragg scattering in the
SC-state with coexisting nematic order. We have made a series of calculations for the ARPES autocorrelation
${\bar C}_{\varsigma}({\bf q},\omega)$, and the results show that the breaking of the $C_{4}$ rotation
symmetry also occurs in the ARPES autocorrelation pattern (then the QSI pattern). To see this exotic feature
more clearly, we plot the intensity map of ${\bar C}_{\varsigma}({\bf q},\omega)$ in a $[q_{x},q_{y}]$
plane for the binding-energy $\omega=94$ meV at $\delta=0.06$ with $T=0.002J$ and the optimal strength of the
electronic nematicity $\varsigma=0.022$ in Fig. \ref{autocorrelation-maps}a, where the positions of the Bragg
peaks ${\bf Q}^{\rm (B)}_{x}$ and ${\bf Q}^{\rm (B)}_{y}$ are labelled by circles, while ${\bf q}_{1}$,
${\bf q}_{2}$, ${\bf q}_{3}$, ${\bf q}_{4}$, ${\bf q}_{5}$, ${\bf q}_{6}$, and ${\bf q}_{7}$ indicate
different scattering wave vectors, however, the scattering wave vectors ${\bf q}_{1}$, ${\bf q}_{4}$, and
${\bf q}_{5}$ and the respective symmetry-corresponding partners occur with unequal amplitudes as in the case
shown in Fig. \ref{EFS-map}. For a comparison, the corresponding experimental result \cite{Fujita19} of the
QSI pattern observed from Bi$_{2}$Sr$_{2}$CaCu$_{2}$O$_{8+\delta}$ at doping $\delta=0.06$ for the bind-energy
$\omega=94$ meV is also shown in Fig. \ref{autocorrelation-maps}b. Obviously, the experimental result
\cite{Fujita19} of the momentum-space structure of the QSI pattern is qualitatively reproduced, where two
distinct classes of the broken-symmetry states in Fig. \ref{autocorrelation-maps}a have emerged as: (i) The
Bragg peaks at the wave vectors ${\bf Q}^{\rm (B)}_{x}=[\pm 2\pi,0]$ along the $\hat{x}$ axis and
${\bf Q}^{\rm (B)}_{y}=[0,\pm 2\pi]$ along the $\hat{y}$ axis, which are the signature of the nematic-order
state with the broken $C_{4}$ rotation symmetry, since the intensity of the peak at the Bragg wave vector
${\bf Q}^{\rm (B)}_{x}$ is different from that at the Bragg wave vector ${\bf Q}^{\rm (B)}_{y}$. To see this
difference more clearly, we plot ${\bar C}_{\varsigma}({\bf q},\omega)$ as a function of momentum at around
(a) the Bragg wave vector ${\bf Q}^{\rm (B)}_{x}$ along the $q_{x}$ direction and (b) at around the Bragg
wave vector ${\bf Q}^{\rm (B)}_{y}$ along the $q_{y}$ direction in the binding-energy $\omega=94$ meV for
$\delta=0.06$ with $T=0.002J$ and the optimal strength of the electronic nematicity $\varsigma=0.022$ in
Fig. \ref{Bragg-peak}, where when the momentum is turned away from the Bragg wave vector
${\bf Q}^{\rm (B)}_{x}$ $({\bf Q}^{\rm (B)}_{y})$, the distinct peak at the Bragg wave vector
${\bf Q}^{\rm (B)}_{x}$ $({\bf Q}^{\rm (B)}_{y})$ is suppressed rapidly, and then eventually disappears.
More importantly, the intensity of the peak at the Bragg wave vector ${\bf Q}^{\rm (B)}_{x}$ is
higher than that at the Bragg wave vector ${\bf Q}^{\rm (B)}_{y}$, in qualitative agreement with the
experimental observation from the STS measurements \cite{Lawler10,Fujita14,Zheng17,Fujita19}.
Furthermore, ${\bar C}_{\varsigma}({\bf q},\omega)$ as a function of momentum at around the Bragg wave
vector ${\bf Q}^{\rm (B)}_{x}$ along the $q_{y}$ direction and at around the Bragg wave vector
${\bf Q}^{\rm (B)}_{y}$ along the $q_{x}$ direction have been also calculated, and the similar results as
shown in Fig. \ref{Bragg-peak} are obtained. This difference of the intensities of the Bragg peak between
the Bragg wave vectors ${\bf Q}^{\rm (B)}_{x}$ and ${\bf Q}^{\rm (B)}_{y}$ shows the inequivalence on the
average of the electronic structure at the two Bragg scattering sites ${\bf Q}^{\rm (B)}_{x}$ and
${\bf Q}^{\rm (B)}_{y}$, and therefore verifies the $C_{4}$ rotation symmetry-breaking in the presence of
the nematic order; (ii) The peaks at the scattering wave vectors ${\bf q}_{1}$, ${\bf q}_{4}$, and
${\bf q}_{5}$, which are the signatures of the multiple ordered states with broken both rotation and
translation symmetries, while the peaks at the scattering wave vectors ${\bf q}_{2}$, ${\bf q}_{3}$,
${\bf q}_{6}$, and ${\bf q}_{7}$, which are the signatures of the multiple ordered states with broken
translation symmetry. These results are also expected from the {\it octet} scattering model shown in
Fig. \ref{EFS-map} with broken $C_{4}$ rotation symmetry.

\begin{figure}[h!]
\centering
\includegraphics[scale=0.75]{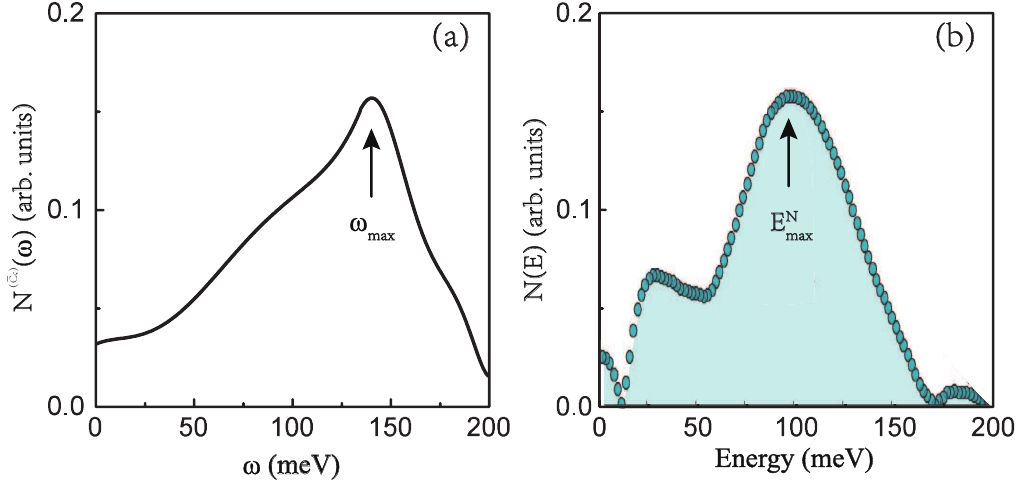}
\caption{(Color online) (a) The energy evolution of the order parameter of the nematic-order state at
$\delta=0.06$ with $T=0.002J$ in the optimal strength of the electronic nematicity $\varsigma=0.022$
for $t/J=3$, $t'/t=1/3$, and $J=100$ meV, where the arrow indicates the positions of the peak. (b) The
corresponding experimental result observed on Bi$_{2}$Sr$_{2}$CaCu$_{2}$O$_{8+\delta}$ at doping
$\delta=0.06$ taken from Ref. \onlinecite{Fujita19}. \label{order-parameter}}
\end{figure}

Following the common practice, the order parameter of the electronic nematicity can be properly represented as
a traceless symmetric tensor \cite{Fradkin10}. In the present case in which the nematic order is associated
with the breaking of the $C_{4}$ rotation symmetry in the underlying square-lattice CuO$_{2}$ plane, the order
parameter of the electronic nematicity is defined by principle as,
\begin{eqnarray*}\label{nematic-order-parameter-1}
N^{({\bar C}_{\varsigma})}_{0}(\omega)={{\bar C}_{\varsigma}({\bf Q}^{\rm (B)}_{x},\omega)
-{\bar C}_{\varsigma}({\bf Q}^{\rm (B)}_{y},\omega)\over {\bar C}_{\varsigma}({\bf Q}^{\rm (B)}_{x},\omega)
+{\bar C}_{\varsigma}({\bf Q}^{\rm (B)}_{y},\omega)}.
\end{eqnarray*}
However, as we have mentioned in Sec. \ref{Formalism}, the calculation for the normal and anomalous
self-energies in Eq. (\ref{TOT-SE}) is performed numerically on a $120\times 120$ lattice in momentum space,
with the infinitesimal $i0_{+}\rightarrow i\Gamma$ replaced by a small damping $\Gamma=0.05J$, which leads
to that the peak weight of the ARPES autocorrelation ${\bar C}_{\varsigma}({\bf q},\omega)$ in
Eq. (\ref{ARPES-autocorrelation}) at the Bragg wave vector ${\bf Q}^{\rm (B)}_{x}$ $[{\bf Q}^{\rm (B)}_{y}]$
spreads on the extremely small area $\{{\bf Q}^{\rm (B)}_{x}\}$ $[\{{\bf Q}^{\rm (B)}_{y}\}]$ around the
${\bf Q}^{\rm (B)}_{x}$ $[{\bf Q}^{\rm (B)}_{y}]$ point as shown in Fig. \ref{Bragg-peak}a
[Fig. \ref{Bragg-peak}b]. In particular, the summation of these spread weights around this extremely small
area $\{{\bf Q}^{\rm (B)}_{x}\}$ $[\{{\bf Q}^{\rm (B)}_{y}\}]$ is less affected by the calculation for a
finite lattice. In this case, a more appropriate order parameter of the electronic nematicity can be defined
as,
\begin{eqnarray}\label{nematic-order-parameter}
N^{({\bar C}_{\varsigma})}(\omega)={{\bar C}^{(x)}_{\varsigma}(\omega)-{\bar C}^{(y)}_{\varsigma}(\omega)\over
{\bar C}^{(x)}_{\varsigma}(\omega)+{\bar C}^{(y)}_{\varsigma}(\omega)},
\end{eqnarray}
for the reduction of the size effect in the finite-lattice calculation, where
${\bar C}^{(x)}_{\varsigma}(\omega)=(1/N)\sum_{{\bf q}\in \{{\bf Q}^{\rm (B)}_{x}\}}
{\bar C}_{\varsigma}({\bf q},\omega)$ and ${\bar C}^{(y)}_{\varsigma}(\omega)=(1/N)
\sum_{{\bf q}\in \{{\bf Q}^{\rm (B)}_{y}\}}{\bar C}_{\varsigma}({\bf q},\omega)$, with the summation
${\bf q}\in \{{\bf Q}^{\rm (B)}_{x}\}$ $[{\bf q}\in \{{\bf Q}^{\rm (B)}_{y}\}]$ that is restricted to the
extremely small area $\{{\bf Q}^{\rm (B)}_{x}\}$ $[\{{\bf Q}^{\rm (B)}_{y}\}]$ around
${\bf Q}^{\rm (B)}_{x}$ $[{\bf Q}^{\rm (B)}_{y}]$. This order parameter
$N^{({\bar C}_{\varsigma})}(\omega)$ is identically zero in the state without the $C_{4}$ rotation
symmetry-breaking, and non-zero in the nematic-order state with broken $C_{4}$ rotation symmetry. We have
carried out a series of calculation for the order parameter $N^{({\bar C}_{\varsigma})}(\omega)$, and the
result shows that order parameter $N^{({\bar C}_{\varsigma})}(\omega)|_{\omega=94 {\rm meV}}\sim 0.1\neq 0$
in the binding-energy $\omega=94$ meV. This non-zero order parameter further confirms the breaking of
the $C_{4}$ rotation symmetry-breaking in the ARPES autocorrelation (then QSI) shown in
Fig. \ref{autocorrelation-maps} in the presence of the electronic nematicity. Moreover, this order
parameter $N^{({\bar C}_{\varsigma})}(\omega)$ is strongly energy dependent. To see this strong energy
dependence of the order parameter more clearly, we plot $N^{({\bar C}_{\varsigma})}(\omega)$ as a function
of binding-energy at $\delta=0.06$ with $T=0.002J$ in the optimal strength of the electronic nematicity
$\varsigma=0.022$ in Fig. \ref{order-parameter}a. For a comparison, the corresponding experimental result
\cite{Fujita19} of the energy dependence of the order parameter of the electronic nematicity observed from
Bi$_{2}$Sr$_{2}$CaCu$_{2}$O$_{8+\delta}$ at doping $\delta=0.06$ is also shown in Fig. \ref{order-parameter}b.
Obviously, the global feature of the obtained energy dependence of the order parameter exhibits
qualitatively resemblance to the corresponding experimental data \cite{Fujita19}. This order parameter
$N^{({\bar C}_{\varsigma})}(\omega)$ starts to grow continuously with the increase of binding-energy in the
lower energy range, and achieves its maximum in the energy $\omega_{\rm max}\sim 1.4J$, then decrease
rapidly with the increase of binding-energy in the higher energy range. In particular, this anticipated
characteristic energy $\omega_{\rm max}\sim 1.4J=140$ meV is not too far from $E^{(N)}_{\rm max}\sim 94$
meV observed from the QSI patterns \cite{Fujita19} in Bi$_{2}$Sr$_{2}$CaCu$_{2}$O$_{8+\delta}$ at doping
$\delta=0.06$. In this case, the electronic nematicity verified using the order parameter
$N^{({\bar C}_{\varsigma})}(\omega)$ indeed derives primarily from the inequivalence of the electronic
structure at the two Bragg scattering sites ${\bf Q}^{\rm (B)}_{x}$ and ${\bf Q}^{\rm (B)}_{y}$. However,
it should be emphasized that the magnitude of the characteristic energy $\omega_{\rm max}$ is also
dependent on the strength of the electronic nematicity. In particular, for the strength of the electronic
nematicity $\varsigma=0.006$, the corresponding $\omega_{\rm max}\sim 0.936J=93.6$ meV, which is well
consistent with the experimental result \cite{Fujita19} of $E^{(N)}_{\rm max}\sim 94$ meV.

Experimentally, by measuring tunneling conductance of Bi$_{2}$Sr$_{2}$CaCu$_{2}$O$_{8+\delta}$ over a
large field of view, performing a FT, and analyzing data from distinct regions of momentum space
\cite{Fujita19}, the pseudogap energy $\Delta^{*}$, the charge-order characteristic energy
$\Delta^{\rm (co)}_{\rm max}$, and the electronic nematicity characteristic energy $\omega_{\rm max}$
have been identified, where measured on the samples whose doping spans the pseudogap regime,
$\omega_{\rm max}$, $\Delta^{\rm (co)}_{\rm max}$, and $\Delta^{*}$ are, within the experimental error,
identical \cite{Fujita19}. These experimental observations
therefore strongly suggest that the nematic and charge orders are tied to the pseudogap. On the
theoretical hand, we \cite{Cao21} have studied very recently the doping dependence of the electronic
nematicity characteristic energy in cuprate superconductors, and the obtained result is fully
consistent with the corresponding result \cite{Fujita19} observed on
Bi$_{2}$Sr$_{2}$CaCu$_{2}$O$_{8+\delta}$. In particular, the obtained doping dependence of the
electronic nematicity characteristic energy is also in good agreement with the corresponding doping
dependence of the pseudogap energy $\Delta^{*}$ over a wide range of doping \cite{Fujita19}. The
pseudogap in the framework of the kinetic energy driven superconductivity originates from the electron
self-energy \cite{Feng12,Feng15a}, and then it can be identified as being a region of the electron
self-energy effect in which the pseudogap suppresses the spectral weight of the quasiparticle
excitation spectrum. The present study together with the study in Ref. \onlinecite{Cao21} thus show that
the pseudogap regime harbours diverse manifestations of the electronically ordered phases, and then
a characteristic feature in the complicated phase diagram of cuprate superconductors is the
coexistence and intertwinement of different ordered states and superconductivity. Moreover, the
nematic-order state strength dependence of the characteristic energy $\omega_{\rm max}$ has been
also discussed \cite{Cao21}, and the result shows that the nematic-order state strength dependence
of the characteristic energy $\omega_{\rm max}$ coincides with the corresponding nematic-order state
strength dependence of the enhancement of $T_{\rm c}$, suggesting a possible connection between the
electronic nematicity characteristic energy and the enhancement of the superconductivity.

\section{Summary and discussions}\label{conclusions}

Within the framework of the kinetic-energy-driven superconductivity, we have investigated the intertwinement
of the electronic nematicity with superconductivity in cuprate superconductors. The obtained phase diagram
shows that the magnitude of the optimized $T_{\rm c}$ is found to increase gradually with the increase of
the strength of the electronic nematicity in the weak strength region, and reaches its maximum in the optimal
strength of the electronic nematicity, then monotonically decreases with the increase of the strength of the
electronic nematicity in the strong strength region, in a striking similar to the dome-like shape doping
dependence of $T_{\rm c}$. This dome-like shape nematic-order strength dependence of $T_{\rm c}$ therefore
indicates that superconductivity in cuprate superconductors is enhanced by the electronic nematicity. Our
results also show that the electronic nematicity has high impacts on the electronic structure, where (i)
the original EFS contour with the $C_{4}$ rotation symmetry is broken up into that with a residual $C_{2}$
rotation symmetry, leading to the EFS anisotropy. However, this EFS contour with the $C_{2}$ rotation
symmetry still shrinks down to form the disconnected Fermi arcs with the most spectral weight that
accommodates at around the tips of the Fermi arcs; (ii) although the tips of the Fermi arcs connected by the
scattering wave vectors ${\bf q}_{i} $ still construct an {\it octet} scattering model, for the partial
quasiparticle scattering processes with the corresponding scattering wave vectors ${\bf q}_{1}$,
${\bf q}_{4}$, and ${\bf q}_{5}$, the amplitudes of the scattering wave vectors are respectively inequivalent
to their symmetry-corresponding partners, leading to the rotation symmetry-breaking of the {\it octet}
scattering model; (iii) the line-shape of the energy distribution curve is inequivalent between the antinodal
region near the X point of BZ and the antinodal region near the Y point, leading to the electronic structure
anisotropy. Moreover, these anisotropic features of the electronic structure obtained from the single-particle
excitation spectrum are also confirmed via the results of QSI described in terms of the ARPES autocorrelation,
where the breaking of the $C_{4}$ rotation symmetry is verified by the inequivalence on the average of the
electronic structure at the two Bragg scattering sites. The theoretical results also indicate that the order
parameter of the electronic nematicity achieves its maximum in the characteristic energy $\omega_{\rm max}$,
however, when the energy is tuned away from this characteristic energy $\omega_{\rm max}$, the order parameter
of the electronic nematicity decreases heavily, and then eventually disappears.

\section*{Acknowledgements}

The authors would like to thank Dr. Yingping Mou for helpful discussions.
ZC and SF are supported by the National Key Research and Development Program of China, and the National
Natural Science Foundation of China (NSFC) under Grant Nos. 11974051 and 11734002. HG is supported by NSFC
under Grant Nos. 11774019 and 12074022, and the Fundamental Research Funds for the Central Universities
and HPC resources at Beihang University.

%\vskip 0.5cm

\begin{appendix}

\section{Full electron diagonal and off-diagonal propagators in superconducting-state with coexisting
nematic order} \label{nematic-order-propagators}

In this Appendix, our main goal is to generalize the theoretical framework of the kinetic-energy-driven
superconductivity from the previous case without the rotation symmetry-breaking to the present case with
broken rotation symmetry, and then derive the full electron diagonal and off-diagonal propagators
$G_{\varsigma}({\bf k},\omega)$ and $\Im^{\dagger}_{\varsigma}({\bf k},\omega)$ in Eq. (\ref{EGF}) of the
main text. The $t$-$J$ model (\ref{tJ-model}) in the fermion-spin representation can be expressed as
\cite{Feng9404,Feng15,Feng0306,Feng12,Feng15a},
\begin{eqnarray}\label{CSS-tJ-model}
H&=&\sum_{<l\hat{\eta}>}t_{\hat{\eta}}(h^{\dagger}_{l+\hat{\eta}\uparrow}h_{l\uparrow}S^{+}_{l}
S^{-}_{l+\hat{\eta}}+h^{\dagger}_{l+\hat{\eta}\downarrow}h_{l\downarrow}S^{-}_{l}S^{+}_{l+\hat{\eta}})
\nonumber\\
&-&\sum_{<l\hat{\tau}>}t'_{\hat{\tau}}(h^{\dagger}_{l+\hat{\tau}\uparrow}h_{l\uparrow}S^{+}_{l}
S^{-}_{l+\hat{\tau}}+ h^{\dagger}_{l+\hat{\tau}\downarrow}h_{l\downarrow}S^{-}_{l}S^{+}_{l+\hat{\tau}})
\nonumber\\
&-&\mu_{\rm h}\sum_{l\sigma}h^{\dagger}_{l\sigma}h_{l\sigma}+\sum_{<l\hat{\eta}>}
J^{(\hat{\eta})}_{\rm eff}{\bf S}_{l}\cdot {\bf S}_{l+\hat{\eta}},
\end{eqnarray}
with the charge-carrier chemical potential $\mu_{\rm h}$,
$J_{\rm eff}^{(\hat{\eta})}=(1-\delta)^{2}J_{\hat{\eta}}$, and the doping concentration
$\delta=\langle h^{\dagger}_{l\sigma}h_{l\sigma}\rangle$.

\subsection{Mean-field theory}\label{Mean-field-theory}

In the mean-field (MF) level, this $t$-$J$ model (\ref{CSS-tJ-model}) in the fermion-spin
representation can be decoupled as \cite{Feng9404,Feng15,Feng0306,Feng12,Feng15a},
\begin{subequations}\label{MF-t-J-model}
\begin{eqnarray}
H_{\rm MF}&=&H^{(\varsigma)}_{\rm t}+H^{(\varsigma)}_{\rm J}+H^{(\varsigma)}_{0}, \\
H^{(\varsigma)}_{\rm t}&=&\sum_{<l\hat{\eta}>\sigma}t_{\hat{\eta}}\chi_{1\hat{\eta}}
h^{\dagger}_{l+\hat{\eta}\sigma}h_{l\sigma}-\chi_{2}t'\sum_{<l\hat{\tau}>\sigma}
h^{\dagger}_{l+\hat{\tau}\sigma}h_{l\sigma}\nonumber\\
&-&\mu_{\rm h}\sum_{l\sigma}h^{\dagger}_{l\sigma}h_{l\sigma}, ~~~~~~~~\label{MF-t-term}\\
H^{(\varsigma)}_{\rm J}&=& {1\over 2}\sum_{<l\hat{\eta}>}J^{(\hat{\eta})}_{\rm eff}
[\epsilon_{\hat{\eta}}(S^{+}_{l}S^{-}_{l+\hat{\eta}}+S^{-}_{l}S^{+}_{l+\hat{\eta}})
+2S^{\rm z}_{l}S^{\rm z}_{l+\hat{\eta}}]\nonumber\\
&-& t'\phi_{2}\sum_{<l\hat{\tau}>}(S^{+}_{l}S^{-}_{l+\hat{\tau}}+S^{-}_{l}
S^{+}_{l+\hat{\tau}}),~~~~~~~~~\label{MF-J-term}
\end{eqnarray}
\end{subequations}
where $H^{(\varsigma)}_{0}=-4N(t_{\hat{x}}\phi_{1{\hat{x}}}\chi_{1{\hat{x}}}+t_{\hat{y}}\phi_{1{\hat{y}}}
\chi_{1{\hat{y}}})+8Nt'\phi_{2}\chi_{2}$,
$\epsilon_{\hat{\eta}}=1+2t_{\hat{\eta}}\phi_{1\hat{\eta}}/J_{\rm eff}^{(\hat{\eta})}$ with
$\hat{\eta}=\hat{x},~\hat{y}$, the charge-carrier's particle-hole parameters
$\phi_{1\hat{\eta}}=\langle h^{\dagger}_{l\sigma}h_{l+\hat{\eta}\sigma}\rangle$ and
$\phi_{2}=\langle h^{\dagger}_{l\sigma}h_{l+\hat{\tau}\sigma}\rangle$, and the spin correlation
functions $\chi_{1\hat{\eta}}=\langle S_{l}^{+}S_{l+\hat{\eta}}^{-}\rangle$ and
$\chi_{2}=\langle S_{l}^{+}S_{l+\hat{\tau}}^{-}\rangle$, while the hoping amplitudes $t_{\hat{x}}$ and
$t_{\hat{y}}$ have been given explicitly in Eq. (\ref{NO-parameter}) of the main text.

From the charge-carrier part (\ref{MF-t-term}), it is straightforward to obtain the MF charge-carrier
propagator as \cite{Feng9404,Feng15,Feng0306,Feng12,Feng15a},
\begin{eqnarray}\label{MFHGF}
g^{(0)}_{\varsigma}({\bf k},\omega)={1\over \omega-\xi^{(\varsigma)}_{\bf k}},
\end{eqnarray}
where the MF charge-carrier orthorhombic energy dispersion is evaluated directly from
Eq. (\ref{MF-t-term}), and can be expressed explicitly as,
\begin{eqnarray}\label{MFCCS}
\xi^{(\varsigma)}_{\bf k} &=& 4t[(1-\varsigma)\chi_{1\hat{x}}\gamma_{{\bf k}_{x}}+(1+\varsigma)
\chi_{1\hat{y}}\gamma_{{\bf k}_{y}}]\nonumber\\
&-& 4t'\chi_{2}\gamma_{\bf k}'-\mu_{\rm h}. ~~~~~
\end{eqnarray}

On the other hand, in the doped regime without an AFLRO, i.e., $\langle S^{\rm z}_{l}\rangle =0$, the MF
spin propagator can be derived in terms of the Kondo-Yamaji decoupling scheme \cite{Kondo72}, which is a
stage one-step further than the Tyablikov's decoupling scheme \cite{Tyablikov67}. However, in the MF level,
the spin part of the $t$-$J$ model in Eq. (\ref{MF-J-term}) is anisotropic Heisenberg model, it thus needs
two spin propagators
$D_{\varsigma}(l-l',t-t')=-i\theta(t-t')\langle [S^{+}_{l}(t),S^{-}_{l'}(t')]\rangle
=\langle\langle S^{+}_{l}(t);S^{-}_{l'}(t')\rangle\rangle$ and
$D_{\varsigma{\rm z}}(l-l',t-t')=-i\theta(t-t')\langle [S^{\rm z}_{l}(t),S^{\rm z}_{l'}(t')]\rangle
=\langle\langle S^{\rm z}_{l}(t);S^{\rm z}_{l'}(t')\rangle\rangle$
to give a proper description of the nature of the spin excitation
\cite{Feng9404,Feng15,Feng0306,Feng12,Feng15a}. Following our previous discussions in the case without the
rotation symmetry-breaking \cite{Feng9404,Feng15,Feng0306,Feng12,Feng15a}, the MF spin propagators
$D^{(0)}_{\varsigma}({\bf k},\omega)$ and $D^{(0)}_{\rm \varsigma z}({\bf k},\omega)$ in the present case
with broken rotation symmetry can be derived as,
\begin{subequations}\label{TWO-MFSGF}
\begin{eqnarray}
D^{(0)}_{\varsigma}({\bf k},\omega)&=&{B^{(\varsigma)}_{\bf k}\over 2\omega^{(\varsigma)}_{\bf k}}
\left ({1\over \omega-\omega^{(\varsigma)}_{\bf k}}
-{1\over\omega+\omega^{(\varsigma)}_{\bf k}}\right ),\label{MFSGF}\\
D^{(0)}_{\rm \varsigma z}({\bf k},\omega)&=& {B^{(\varsigma)}_{{\rm z}{\bf k}}\over
2\omega^{(\varsigma)}_{{\rm z}{\bf k}}}\left ({1\over\omega-\omega^{(\varsigma)}_{{\rm z}{\bf k}}}
-{1\over\omega+\omega^{(\varsigma)}_{{\rm z}{\bf k}}}\right ),~~~~
\label{MFSGFZ}
\end{eqnarray}
\end{subequations}
where the weight functions $B^{(\varsigma)}_{{\bf{k}}}$ and $B^{(\varsigma)}_{{\rm z}{\bf k}}$ are given by,
\begin{widetext}
\begin{subequations}
\begin{eqnarray}
B^{(\varsigma)}_{{\bf{k}}} &=& \lambda_{1\hat{x}}[ 2\chi^{\rm z}_{1\hat{x}} (\epsilon_{\hat{x}}
\gamma_{{\bf k}_{x}}-{1\over 2}) +\chi_{1\hat{x}}( \gamma_{{\bf k}_{x}}-{1\over 2}\epsilon_{\hat{x}})]
+\lambda_{1\hat{y}}[ 2\chi^{\rm z}_{1\hat{y}}(\epsilon_{\hat{y}}\gamma_{{\bf k}_{y}}-{1\over 2})
+\chi_{1\hat{y}} (\gamma_{{\bf k}_{y}}-{1\over 2}\epsilon_{\hat{y}})]
-\lambda_{2}(2\chi^{\rm z}_{2}\gamma_{\bf{k}}'-\chi_{2}),~~~~~~\\
B^{(\varsigma)}_{{\rm z}{\bf k}}&=&\lambda_{1\hat{x}}\epsilon_{\hat{x}}\chi_{1\hat{x}}(
\gamma_{{\bf k}_{x}}-{1\over 2})+\lambda_{1\hat{y}}\epsilon_{\hat{y}}\chi_{1\hat{y}}
( \gamma_{{\bf k}_{y}}-{1\over 2})-\lambda_{2}\chi_{2}(\gamma_{\bf{k}}'-1),
\end{eqnarray}
\end{subequations}
%\end{widetext}
and the MF spin orthorhombic excitation spectra $\omega^{(\varsigma)}_{{\rm z}{\bf k}}$ and
$\omega^{(\varsigma)}_{\bf k}$ are obtained explicitly as,
%\begin{widetext}
\begin{subequations}\label{TWO-MFSES}
\begin{eqnarray}
(\omega^{(\varsigma)}_{{\rm z}{\bf k}})^{2} &=&\epsilon_{\hat{x}}\lambda^{2}_{1\hat{x}}\left (
\epsilon_{\hat{x}}A_{1\hat{x}}-{1\over Z}\alpha\chi^{\rm z}_{1\hat{x}}-\alpha\chi_{1\hat{x}}
\gamma_{{\bf k}_{x}}\right ) (1-\gamma_{{\bf k}_{x}})+\epsilon_{\hat{y}}
\lambda^{2}_{1\hat{y}} \left ( \epsilon_{\hat{y}}A_{1\hat{y}}-{1\over Z}\alpha
\chi^{\rm z}_{1\hat{y}}-\alpha\chi_{1\hat{y}}\gamma_{{\bf k}_{y}}\right )
(1-\gamma_{{\bf k}_{y}})\nonumber\\
&+&\lambda^{2}_{2}A_{3}(1-\gamma_{\bf{k}}')
+\lambda_{1\hat{x}}\lambda_{2}[\alpha\epsilon_{\hat{x}}C_{3\hat{x}}(\gamma_{{\bf k}_{x}}
-1)+\alpha(\chi_{2}\gamma_{{\bf k}_{x}}-\epsilon_{\hat{x}}C_{3\hat{x}})(1-\gamma_{\bf{k}}')]
+\lambda_{1\hat{y}}\lambda_{2}[\alpha\epsilon_{\hat{y}}C_{3\hat{y}}(\gamma_{{\bf k}_{y}}-1)
\nonumber\\
&+&\alpha(\chi_{2}\gamma_{{\bf k}_{y}}-\epsilon_{\hat{y}}C_{3\hat{y}})(1-\gamma_{\bf{k}}')]
+\lambda_{1\hat{x}}\lambda_{1\hat{y}}\left [ \alpha(\epsilon_{\hat{x}}\chi_{1\hat{x}}
+\epsilon_{\hat{y}}\chi_{1\hat{y}})\gamma_{{\bf k}_{x}}\gamma_{{\bf k}_{y}}-{\alpha\over 2}
\epsilon_{\hat{y}}(\epsilon_{\hat{x}}C_{1\hat{x}\hat{y}}+\chi_{1\hat{y}})\gamma_{{\bf k}_{x}}
\right .\nonumber\\
&-&\left.{\alpha\over 2}\epsilon_{\hat{x}}(\epsilon_{\hat{y}}C_{1\hat{x}\hat{y}}+\chi_{1\hat{x}})
\gamma_{{\bf k}_{y}}+{\alpha\over 2}\epsilon_{\hat{x}}\epsilon_{\hat{y}}C_{1\hat{x}\hat{y}}
\right ], \label{MFSESZ}
\end{eqnarray}
\begin{eqnarray}
(\omega^{(\varsigma)}_{\bf k})^{2}&=&\lambda^{2}_{1\hat{x}}\left [ {1\over 2}\epsilon_{\hat{x}}
\left ( A_{1\hat{x}}-{1\over 2}\alpha\chi^{\rm z}_{1\hat{x}}-\alpha\chi_{1\hat{x}}
\gamma_{{\bf k}_{x}}\right ) (\epsilon_{\hat{x}}-\gamma_{{\bf k}_{x}})+\left (
A_{2\hat{x}}-{1\over 2Z}\alpha\epsilon_{\hat{x}}\chi_{1\hat{x}}-\alpha\epsilon_{\hat{x}}
\chi^{\rm z}_{1\hat{x}}\gamma_{{\bf k}_{x}}\right ) (1-\epsilon_{\hat{x}}
\gamma_{{\bf k}_{x}})\right ] \nonumber\\
&+&\lambda^{2}_{1\hat{y}}\left [ {1\over 2}\epsilon_{\hat{y}}\left( A_{1\hat{y}}
-{1\over 2}\alpha\chi^{\rm z}_{1\hat{y}}-\alpha\chi_{1\hat{y}}\gamma_{{\bf k}_{y}}
\right ) (\epsilon_{\hat{y}}-\gamma_{{\bf k}_{y}})+\left ( A_{2\hat{y}}-{1\over 2Z}
\alpha\epsilon_{\hat{y}}\chi_{1\hat{y}}-\alpha\epsilon_{\hat{y}}\chi^{\rm z}_{1\hat{y}}
\gamma_{{\bf k}_{y}}\right ) (1-\epsilon_{\hat{y}}\gamma_{{\bf k}_{y}})\right ]
\nonumber\\
&+&\lambda^{2}_{2}\left [ \alpha \left(\chi^{\rm z}_{2}\gamma_{\bf{k}}'-{3\over 2Z}
\chi_{2}\right ) \gamma_{\bf{k}}'+{1\over 2}\left( A_{3}-{1\over 2}\alpha\chi^{\rm z}_{2}
\right )\right ]+\lambda_{1\hat{x}}\lambda_{2}\left [ \alpha\chi^{\rm z}_{1\hat{x}}
(1-\epsilon_{\hat{x}}\gamma_{{\bf k}_{x}})\gamma_{\bf{k}}'+{1\over 2}\alpha(\chi_{1\hat{x}}
\gamma_{\bf{k}}'-C_{3\hat{x}})(\epsilon_{\hat{x}}-\gamma_{{\bf k}_{x}})\right . \nonumber\\
&+&\left . \alpha\gamma_{\bf{k}}'(C^{\rm z}_{3\hat{x}}-\epsilon_{\hat{x}}\chi^{\rm z}_{2}
\gamma_{{\bf k}_{x}})-{1\over 2}\alpha\epsilon_{\hat{x}}(C_{3\hat{x}}-\chi_{2}\gamma_{{\bf k}_{x}})
\right] +\lambda_{1\hat{y}}\lambda_{2}\left [\alpha\chi^{\rm z}_{1\hat{y}}(1-\epsilon_{\hat{y}}
\gamma_{{\bf k}_{y}})\gamma_{\bf{k}}'+{1\over 2}\alpha(\chi_{1\hat{y}}\gamma_{\bf{k}}'
-C_{3\hat{y}})(\epsilon_{\hat{y}}-\gamma_{{\bf k}_{y}})\right .\nonumber\\
&+&\left . \alpha\gamma_{\bf{k}}'(C^{\rm z}_{3\hat{y}}-\epsilon_{\hat{y}}\chi^{\rm z}_{2}
\gamma_{{\bf k}_{y}})-{1\over 2}\alpha\epsilon_{\hat{y}}(C_{3\hat{y}}-\chi_{2}
\gamma_{{\bf k}_{y}}) \right]+\lambda_{1\hat{x}}\lambda_{1\hat{y}}\left \{ \alpha
\left[\epsilon_{\hat{x}}\epsilon_{\hat{y}}(\chi^{\rm z}_{1\hat{x}}+\chi^{\rm z}_{1\hat{y}})
+{1\over 2}(\epsilon_{\hat{y}}\chi_{1\hat{x}}-\epsilon_{\hat{x}}\chi_{1\hat{y}})\right ]
\gamma_{{\bf k}_{x}}\gamma_{{\bf k}_{y}}\right.\nonumber\\
&-&{\alpha\over 2}\left[\epsilon_{\hat{x}}(C^{\rm z}_{1\hat{x}\hat{y}}+\chi^{\rm z}_{1\hat{y}})
+{1\over 2}\epsilon_{\hat{y}}(\epsilon_{\hat{x}}\chi_{1\hat{y}}-C_{1\hat{x}\hat{y}})\right ]
\gamma_{{\bf k}_{x}}
-{\alpha\over 2}\left [ \epsilon_{\hat{y}}(C^{\rm z}_{1\hat{x}\hat{y}}
+\chi^{\rm z}_{1\hat{x}})+{1\over 2}\epsilon_{\hat{x}}(\epsilon_{\hat{y}}\chi_{1\hat{x}}
-C_{1\hat{x}\hat{y}})\right ] \gamma_{{\bf k}_{y}}\nonumber\\
&+&\left.{\alpha\over Z}(\epsilon_{\hat{x}}
\epsilon_{\hat{y}}C_{1\hat{x}\hat{y}}+2C^{\rm z}_{1\hat{x}\hat{y}}) \right\}, \label{MFSES}
\end{eqnarray}
\end{subequations}
with $\lambda_{1\hat{x}}=2ZJ^{(\hat{x})}_{\rm{eff}}$,
$\lambda_{1\hat{y}}=2ZJ^{(\hat{y})}_{\rm{eff}}$, $\lambda_{2}=4Z\phi_{2}t'$,
$A_{1\hat{x}}=\alpha C_{1\hat{x}}+(1-\alpha)/(2Z)$,
$A_{1\hat{y}}=\alpha C_{1\hat{y}}+(1-\alpha)/(2Z)$,
$A_{2\hat{x}}=\alpha C^{\rm z}_{1\hat{x}}+(1-\alpha)/(4Z)$,
$A_{2\hat{y}}=\alpha C^{\rm z}_{1\hat{y}}+(1-\alpha)/(4Z)$, $A_{3}=\alpha C_{2}+(1-\alpha)/(2Z)$,
the spin correlation functions
$\chi^{\rm z}_{1\hat{x}}=\left \langle S^{\rm z}_{l}S^{\rm z}_{l+\hat{x}}\right\rangle$,
$\chi^{\rm z}_{1\hat{y}}=\left\langle S^{\rm z}_{l}S^{\rm z}_{l+\hat{y}}\right\rangle$,
$\chi^{\rm z}_{2}=\left\langle S^{\rm z}_{l}S^{\rm z}_{l+\hat{\tau}}\right\rangle$,
$C_{1\hat{x}}=(4/Z^{2})\sum_{\hat{x},\hat{x}'}\left\langle S^{+}_{l+\hat{x}}S^{-}_{l+\hat{x}'}\right\rangle$,
$C_{1\hat{y}}=(4/Z^{2})\sum_{\hat{y},\hat{y}'}\left\langle S^{+}_{l+\hat{y}}S^{-}_{l+\hat{y}'}\right\rangle$,
$C_{1\hat{x}\hat{y}}=(4/Z^{2})\sum_{\hat{x},\hat{y}}\left\langle S^{+}_{l+\hat{x}}S^{-}_{l+\hat{y}}\right\rangle$,
$C^{\rm z}_{1\hat{x}}=(4/Z^{2})\sum_{\hat{x},\hat{x}'}\left\langle S^{\rm z}_{l+\hat{x}}S^{\rm z}_{l+\hat{x}'}\right\rangle$,
$C^{\rm z}_{1\hat{y}}=(4/Z^{2})\sum_{\hat{y},\hat{y}'}\left\langle S^{\rm z}_{l+\hat{y}}S^{\rm z}_{l+\hat{y}'}\right\rangle$,
$C^{\rm z}_{1\hat{x}\hat{y}}=(4/Z^{2})\sum_{\hat{x},\hat{y}}\left\langle S^{\rm z}_{l+\hat{x}}S^{\rm z}_{l+\hat{y}}\right\rangle$,
$C_{2}=(1/Z^{2})\sum_{\hat{\tau},\hat{\tau}'}\left\langle S^{+}_{l+\hat{\tau}}S^{-}_{l+\hat{\tau}'}\right\rangle$,
$C_{3\hat{x}}=(2/Z^{2})\sum_{\hat{x},\hat{\tau}}\left \langle S^{+}_{l+\hat{x}}S^{-}_{l+\hat{\tau}}\right\rangle$,
$C_{3\hat{y}}=(2/Z^{2})\sum_{\hat{y},\hat{\tau}}\left\langle S^{+}_{l+\hat{y}}S^{-}_{l+\hat{\tau}}\right\rangle$,
$C^{\rm z}_{3\hat{x}}=(2/Z^{2})\sum_{\hat{x},\hat{\tau}}\left\langle S^{\rm z}_{l+\hat{x}}S^{\rm z}_{l+\hat{\tau}}\right\rangle$,
$C^{\rm z}_{3\hat{y}}=(2/Z^{2})\sum_{\hat{y},\hat{\tau}}\left\langle S^{\rm z}_{l+\hat{y}}S^{\rm z}_{l+\hat{\tau}}\right\rangle$,
the number of the NN or next NN sites on a square lattice $Z$. In order to fulfill the sum rule
of the correlation function $\left \langle S^{+}_{l}S^{-}_{l}\right\rangle=1/2$ in the case
without an AFLRO, the important decoupling parameter $\alpha$ has been introduced in the above
calculation, which can be regarded as the vertex correction \cite{Kondo72}.
\end{widetext}

\subsection{Charge-carrier normal and anomalous self-energies}\label{Charge-carrier-self-energy}

In the absence of the electronic nematicity, it has been shown that the interaction between the charge carriers
directly from the kinetic energy of the $t$-$J$ model by the exchange of a strongly dispersive spin excitation
generates the charge-carrier pairing state in the particle-particle channel \cite{Feng9404,Feng15,Feng0306,Feng12}.
Following these previous discussions, the equations that are satisfied by the full charge-carrier diagonal and
off-diagonal propagators of the $t$-$J$ model (\ref{CSS-tJ-model}) in the SC-state with coexisting electronic
nematicity can be also obtained as,
\begin{subequations}\label{CCSCES}
\begin{eqnarray}
g_{\varsigma}({\bf k},\omega)&=&g^{(0)}_{\varsigma}({\bf k},\omega)+g^{(0)}_{\varsigma}({\bf k},\omega)
[\Sigma^{(\rm h)}_{\varsigma{\rm ph}}({\bf k},\omega)g_{\varsigma}({\bf k},\omega)\nonumber\\
&-&\Sigma^{(\rm h)}_{\varsigma{\rm pp}}({\bf k},\omega)\Gamma^{\dagger}_{\varsigma}({\bf k},\omega)],
\end{eqnarray}
\begin{eqnarray}
\Gamma^{\dagger}_{\varsigma}({\bf k},\omega)&=& g^{(0)}_{\varsigma}({\bf k},-\omega)
[\Sigma^{(\rm h)}_{\varsigma{\rm ph}}({\bf k},-\omega)\Gamma^{\dagger}_{\varsigma}({\bf k},\omega)
\nonumber\\
&+& \Sigma^{(\rm h)}_{\varsigma{\rm pp}}({\bf k},\omega)g_{\varsigma}({\bf k},\omega)],~~~~
\end{eqnarray}
\end{subequations}
with the charge-carrier normal self-energy $\Sigma^{(\rm h)}_{\varsigma{\rm ph}}({\bf k},\omega)$ in the
particle-hole channel and the charge-carrier anomalous self-energy
$\Sigma^{(\rm h)}_{\varsigma{\rm pp}}({\bf k},\omega)$ in the particle-particle channel, which can be
evaluated in terms of the spin bubble as \cite{Feng9404,Feng15,Feng0306,Feng12},
%\begin{widetext}
\begin{subequations}\label{CCSE}
\begin{eqnarray}
\Sigma^{({\rm h})}_{\varsigma{\rm ph}}({\bf k},i\omega_{n})&=&{1\over N^{2}}\sum_{{\bf p},{\bf p}'}
[\Lambda^{(\varsigma)}_{{\bf p}+{\bf p}'+{\bf k}}]^{2}\nonumber\\
&\times&{1\over\beta}\sum_{ip_{m}}
g_{\varsigma}({{\bf p}+{\bf k}},ip_{m}+i\omega_{n})\Pi_{\varsigma}({\bf p},{\bf p}',ip_{m}), \nonumber\\
~~ \\
\Sigma^{({\rm h})}_{\varsigma{\rm pp}}({\bf k},i\omega_{n})&=&{1\over N^{2}}\sum_{{\bf p},{\bf p}'}
[\Lambda^{(\varsigma)}_{{\bf p}+{\bf p}'+{\bf k}}]^{2}\nonumber\\
&\times& {1\over\beta}\sum_{ip_{m}}
\Gamma^{\dagger}_{\varsigma}({\bf p}+{\bf k},ip_{m}+i\omega_{n})\Pi_{\varsigma}({\bf p},{\bf p}',ip_{m}),
\nonumber\\
\end{eqnarray}
\end{subequations}
%\end{widetext}
where $\omega_{n}$ and $p_{m}$ are the fermionic and bosonic Matsubara frequencies, respectively,
$\Lambda^{(\varsigma)}_{\bf k}=4t[(1-\varsigma)\gamma_{{\bf k}_{x}}+(1+\varsigma)\gamma_{{\bf k}_{y}}]
-4t'\gamma_{\bf{k}}'$, while the spin bubble $\Pi_{\varsigma}({\bf p},{\bf p}',ip_{m})$ is obtained
in terms of the MF spin propagator $D^{(0)}_{\varsigma}({\bf k},\omega)$ in Eq. (\ref{MFSGF}) as,
\begin{eqnarray}
\Pi_{\varsigma}({\bf p},{\bf p}',ip_{m})&=&{1\over\beta}\sum_{ip_{m}'}D^{(0)}_{\varsigma}({\bf p}',ip_{m}')
\nonumber\\
&\times& D^{(0)}_{\varsigma}({\bf p'+p},ip_{m}'+ip_{m}).~~~~~
\end{eqnarray}
The above equations (\ref{CCSCES}) and (\ref{CCSE}) therefore show that the charge-carrier normal and
anomalous self-energies $\Sigma^{({\rm h})}_{\varsigma{\rm ph}}({\bf k},\omega)$ and
$\Sigma^{({\rm h})}_{\varsigma{\rm pp}}({\bf k},\omega)$ and full charge-carrier diagonal and off-diagonal
propagators $g_{\varsigma}({\bf k},\omega)$ and $\Gamma^{\dagger}_{\varsigma}({\bf k},\omega)$ are related
self-consistently. Moreover, $\Sigma^{(\rm h)}_{\varsigma{\rm ph}}({\bf k},\omega)$ is identified as the
momentum and energy dependence of the charge-carrier pair gap, i.e.,
$\Sigma^{({\rm h})}_{\varsigma{\rm pp}}({\bf k},\omega)=\bar{\Delta}^{\rm (h)}_{\varsigma}({\bf k},\omega)
=\bar{\Delta}^{\rm (h)}_{\varsigma\hat{x}}({\bf k},\omega)
-\bar{\Delta}^{\rm (h)}_{\varsigma\hat{y}}({\bf k},\omega)$, while
$\Sigma^{(\rm h)}_{\varsigma{\rm ph}}({\bf k},\omega)$ describes the momentum and energy dependence of the
charge-carrier quasiparticle coherence, and therefore competes with charge-carrier pairing-state.

Although $\Sigma^{({\rm h})}_{\varsigma{\rm pp}}({\bf k},\omega)$ is an even function of energy,
$\Sigma^{({\rm h})}_{\varsigma{\rm ph}}({\bf k},\omega)$ is not. In this case, we can divide
$\Sigma^{({\rm h})}_{\varsigma{\rm ph}}({\bf k},\omega)$ into its symmetric and antisymmetric parts
as: $\Sigma^{({\rm h})}_{\varsigma{\rm ph}}({\bf k},\omega)=\Sigma^{({\rm h})}_{\varsigma{\rm phe}}({\bf k},\omega)
+\omega\Sigma^{({\rm h})}_{\varsigma{\rm pho}}({\bf k},\omega)$, and then both
$\Sigma^{({\rm h})}_{\varsigma{\rm phe}}({\bf k},\omega)$ and $\Sigma^{({\rm h})}_{\varsigma{\rm pho}}({\bf k},\omega)$
are an even function of energy. Moreover, this antisymmetric part
$\Sigma^{({\rm h})}_{\varsigma{\rm pho}}({\bf k},\omega)$ is related directly with the momentum and
energy dependence of the charge-carrier quasiparticle coherent weight as:
$Z^{{\rm (h)}-1}_{\varsigma{\rm F}}({\bf k},\omega)=1-{\rm Re}\Sigma^{(\rm h)}_{\varsigma{\rm pho}}({\bf k},\omega)$.
In this paper, we focus mainly on the low-energy behaviors, and then
$\bar{\Delta}^{\rm (h)}_{\varsigma}({\bf k},\omega)$ and $Z^{{\rm (h)}}_{\varsigma{\rm F}}({\bf k},\omega)$
can be generally discussed in the static limit, i.e.,
\begin{subequations}
\begin{eqnarray}
\bar{\Delta}^{\rm (h)}_{\varsigma}({\bf k})&=&{1\over 2}[\bar{\Delta}^{\rm (h)}_{\varsigma\hat{x}}{\rm cos}k_{x}
- \bar{\Delta}^{\rm (h)}_{\varsigma\hat{y}}{\rm cos}k_{y}],\label{CCPGF}\\
{1\over Z^{\rm (h)}_{\varsigma{\rm F}}({\bf k})}&=&1-{\rm Re}\Sigma^{(\rm h)}_{\varsigma{\rm pho}}({\bf k},\omega=0).
\label{CCQCWW}
\end{eqnarray}
\end{subequations}
Although $Z^{\rm (h)}_{\varsigma{\rm F}}({\bf k})$ still is a function of momentum, the momentum dependence is
unimportant in a qualitative discussion. According to the ARPES experiments \cite{DLFeng00,Ding01}, the momentum
${\bf k}$ in $Z^{\rm (h)}_{\varsigma{\rm F}}({\bf k})$ can be chosen as,
\begin{eqnarray}\label{CCQCW}
Z^{\rm (h)}_{\varsigma{\rm F}}=Z^{\rm (h)}_{\varsigma{\rm F}}({\bf k})\mid_{{\bf k}=[\pi,0]}.
\end{eqnarray}
On the other hand, the charge-carrier pair gap in Eq. (\ref{CCPGF}) can be also expressed explicitly as,
\begin{eqnarray}
\bar{\Delta}_{\varsigma}^{\rm (h)}({\bf k})=\bar{\Delta}_{\varsigma{\rm d}}^{\rm (h)}\gamma^{\rm (d)}_{\bf k}
+\bar{\Delta}_{\varsigma{\rm s}}^{\rm (h)}\gamma^{\rm (s)}_{\bf k}, \label{CCPGF-1}
\end{eqnarray}
with $\gamma^{\rm (d)}_{\bf k}=({\rm cos}k_{x}-{\rm cos}k_{y})/2$,
$\gamma^{\rm (s)}_{\bf k}=({\rm cos}k_{x}+{\rm cos}k_{y})/2$, the d-wave component of the charge-carrier pair
gap parameter $\bar{\Delta}_{\varsigma{\rm d}}^{\rm (h)}=(\bar{\Delta}_{\varsigma\hat{x}}^{\rm (h)}
+\bar{\Delta}_{\varsigma\hat{y}}^{\rm (h)})/2$ and the s-wave component
$\bar{\Delta}_{\varsigma{\rm s}}^{\rm (h)}=(\bar{\Delta}_{\varsigma\hat{x}}^{\rm (h)}
-\bar{\Delta}_{\varsigma\hat{y}}^{\rm (h)})/2$.

Based on the above static-limit approximation, the renormalized charge-carrier diagonal and off-diagonal propagators
can be obtained from Eq. (\ref{CCSCES}) as,
\begin{subequations}\label{BCSHGF}
\begin{eqnarray}
g^{\rm (RMF)}_{\varsigma}({\bf k},\omega)&=&Z^{\rm (h)}_{\varsigma{\rm F}}\left ( {U^{2}_{\varsigma{\rm h}}({\bf k})\over\omega
-E^{\rm (h)}_{\varsigma}({\bf k})}+{V^{2}_{\varsigma{\rm h}}({\bf k})\over\omega+E^{\rm (h)}_{\varsigma}({\bf k})}\right ),
\nonumber\\
\label{BCSHDGF}\\
\Gamma_{\varsigma}^{{\rm (RMF)}\dagger}({\bf k},\omega)&=&-Z^{\rm (h)}_{\varsigma{\rm F}}{\bar{\Delta}^{\rm (h)}_{\varsigma{\rm Z}}({\bf k})
\over 2E^{\rm (h)}_{\varsigma}({\bf k})}\left ( {1\over \omega-E^{\rm (h)}_{\varsigma}({\bf k})} \right .\nonumber\\
&-&\left . {1\over\omega+E^{\rm (h)}_{\varsigma}({\bf k})}
\right ),~~~~~~~~~~\label{BCSHODGF}
\end{eqnarray}
\end{subequations}
with the charge-carrier quasiparticle energy dispersion
$E^{\rm (h)}_{\varsigma}({\bf k})=\sqrt{\bar{\xi}^{(\varsigma)2}_{\bf k}+\mid\bar{\Delta}^{\rm (h)}_{\varsigma{\rm Z}}({\bf k})\mid^{2}}$,
the renormalized MF charge-carrier orthorhombic energy dispersion
$\bar{\xi}^{(\varsigma)}_{{\bf k}}=Z^{\rm (h)}_{\varsigma{\rm F}}\xi^{(\varsigma)}_{\bf k}$,
the renormalized charge-carrier pair gap
$\bar{\Delta}^{\rm (h)}_{\varsigma{\rm Z}}({\bf k})=Z^{\rm (h)}_{\varsigma{\rm F}}\bar{\Delta}^{\rm (h)}_{\varsigma}({\bf k})$, and the
charge-carrier quasiparticle coherence factors,
\begin{subequations}\label{BCSCF}
\begin{eqnarray}
U^{2}_{\varsigma{\rm h}}({\bf k})&=&{1\over 2}\left (1+{\bar{\xi}^{(\varsigma)}_{{\bf k}}\over E^{\rm (h)}_{\varsigma}({\bf k})}\right ),\\
V^{2}_{\varsigma{\rm h}}({\bf k})&=&{1\over 2}\left (1-{\bar{\xi}^{(\varsigma)}_{{\bf k}}\over E^{\rm (h)}_{\varsigma}({\bf k})}\right ).
\end{eqnarray}
\end{subequations}
In particular, the above charge-carrier quasiparticle coherence factors satisfy the constraint
$U^{2}_{\varsigma{\rm h}}({\bf k})+V^{2}_{\varsigma{\rm h}}({\bf k})=1$ for any momentum ${\bf k}$.

Substituting the charge-carrier renormalized diagonal and off-diagonal propagators
in Eq. (\ref{BCSHGF}) and MF spin propagator in Eq. (\ref{MFSGF}) into Eq. (\ref{CCSE}), the charge-carrier normal
self-energy $\Sigma^{({\rm h})}_{\varsigma{\rm ph}}({\bf k},\omega)$ and
the anomalous self-energy $\Sigma^{({\rm h})}_{\varsigma{\rm pp}}({\bf k},\omega)$ now can be evaluated explicitly as,
\begin{widetext}
\begin{subequations}\label{SE1}
\begin{eqnarray}
{\Sigma}^{\rm(h)}_{\varsigma{\rm ph}}({\bf{k}},\omega)&=&{1\over N^{2}}\sum_{{\bf{p}}{\bf{p}'}{\nu}}(-1)^{\nu+1}
\Omega^{(\rm{h})}_{\varsigma{\bf{p}}{\bf{p}'}{\bf{k}}}\left [ U^{2}_{\varsigma\rm{h}}({\bf p}+{\bf k})\left (
{F^{(\rm{h})}_{\varsigma1\nu}({\bf p},{\bf p}',{\bf k})\over\omega+\omega^{(\nu)}_{\varsigma{\bf p}{\bf p}'}
-E^{\rm (h)}_{\varsigma}({\bf p}+{\bf k})}-{F^{(\rm{h})}_{\varsigma2\nu}({\bf p},{\bf p}',{\bf k})\over\omega-
\omega^{(\nu)}_{\varsigma{\bf p}{\bf p}'}-E^{\rm (h)}_{\varsigma}({\bf p}+{\bf k})}\right )\right. \nonumber\\
&+&\left.  V^{2}_{\varsigma{\rm h}}({\bf p}+{\bf k})\left ({F^{(\rm{h})}_{\varsigma1\nu}({\bf p},{\bf p}',{\bf k})
\over\omega-\omega^{(\nu)}_{\varsigma{\bf p}{\bf p}'}+E^{\rm (h)}_{\varsigma}({\bf p}+{\bf k})}
-{F^{(\rm{h})}_{\varsigma2\nu}({\bf p},{\bf p}',{\bf k})\over\omega
+\omega^{(\nu)}_{\varsigma{\bf p}{\bf p}'}+E^{\rm (h)}_{\varsigma}({\bf p}+{\bf k})}\right ) \right ], ~~~~\\
{\Sigma}^{\rm(h)}_{\varsigma{\rm pp}}({\bf{k}},\omega)&=&{1\over N^{2}}\sum_{{\bf{p}}{\bf{p}'}{\nu}}(-1)^{\nu}
\Omega^{(\rm{h})}_{\varsigma{\bf{p}}{\bf{p}'}{\bf{k}}}{\bar{\Delta}^{\rm (h)}_{\varsigma{\rm Z}}({\bf p}+{\bf k})\over
2E^{\rm (h)}_{\varsigma}({\bf p}+{\bf k})}\left [ \left ( {F^{(\rm{h})}_{\varsigma1\nu}({\bf p},{\bf p}',{\bf k})
\over\omega+\omega^{(\nu)}_{\varsigma{\bf p}{\bf p}'}-E^{\rm (h)}_{\varsigma}({\bf p}+{\bf k})}
-{F^{(\rm{h})}_{\varsigma2\nu}({\bf p},{\bf p}',{\bf k})\over\omega-\omega^{(\nu)}_{\varsigma{\bf p}{\bf p}'}
-E^{\rm (h)}_{\varsigma}({\bf p}+{\bf k})}\right ) \right. \nonumber\\
&-& \left. \left ( {F^{(\rm{h})}_{\varsigma1\nu}({\bf p},{\bf p}',{\bf k})\over\omega-\omega^{(\nu)}_{\varsigma{\bf p}{\bf p}'}
+E^{\rm (h)}_{\varsigma}({\bf p}+{\bf k})}-{F^{(\rm{h})}_{\varsigma2\nu}({\bf p},{\bf p}',{\bf k})\over\omega
+\omega^{(\nu)}_{\varsigma{\bf p}{\bf p}'}+E^{\rm (h)}_{\varsigma}({\bf p}+{\bf k})}\right ) \right ],
\end{eqnarray}
\end{subequations}
%\end{widetext}
respectively, with $\nu=1,2$, the kernel function ${\Omega}^{(\rm{h})}_{\varsigma{\bf{p}}{\bf{p}'}{\bf{k}}}
=Z^{\rm (h)}_{\varsigma{\rm F}}[\Lambda^{(\varsigma)}_{{\bf{p}}+{\bf{p}'}+{\bf{k}}}]^{2}B^{(\varsigma)}_{\bf{p}'}
B^{(\varsigma)}_{\bf{p}+\bf{p}'}/(4\omega^{(\varsigma)}_{{\bf p}'}\omega^{(\varsigma)}_{{\bf p}+{\bf p}'})$,
$\omega^{(\nu)}_{\varsigma{\bf p}{\bf p}'}={\omega}^{(\varsigma)}_{\bf{p}+\bf{p}'}
-(-1)^{\nu}{\omega}^{(\varsigma)}_{\bf{p}'}$, and the functions,
%\begin{widetext}
\begin{subequations}
\begin{eqnarray}
F^{(\rm{h})}_{\varsigma1\nu}({\bf p},{\bf p}',{\bf k})&=&n_{\rm{F}}[E^{\rm (h)}_{\varsigma}({\bf p}+{\bf k})]\{
1+n_{\rm B}(\omega^{(\varsigma)}_{\bf{p}'+\bf{p}})+ n_{\rm B}[(-1)^{\nu+1}\omega^{(\varsigma)}_{\bf{p}'}]\}
+n_{\rm B}(\omega^{(\varsigma)}_{\bf{p}'+\bf{p}})n_{\rm B}[(-1)^{\nu+1}\omega^{(\varsigma)}_{\bf{p}'}], \\
F^{(\rm{h})}_{\varsigma2\nu}({\bf{p}},{\bf{p}'},{\bf{k}})&=&\{1-n_{\rm{F}}[E^{\rm (h)}_{\varsigma}({\bf{p}}+{\bf{k}})]\}
\{1+n_{\rm B}(\omega^{(\varsigma)}_{\bf{p}'+\bf{p}})+ n_{\rm B}[(-1)^{\nu+1}\omega^{(\varsigma)}_{\bf{p}'}]\}
+n_{B}(\omega^{(\varsigma)}_{\bf{p}'+\bf{p}})n_{B}[(-1)^{\nu+1}\omega^{(\varsigma)}_{\bf{p}'}],~~~~~~~~
\end{eqnarray}
\end{subequations}
%\end{widetext}
where $n_{B}(\omega)$ and $n_{F}(\omega)$ are the boson and fermion distribution functions,
respectively.

\subsection{Self-consistent equations for determination of charge-carrier order parameters}\label{CSCE}

The above charge-carrier quasiparticle coherent weight $Z^{\rm (h)}_{\varsigma{\rm F}}$ and two components
of the charge carrier pair gap parameter $\bar{\Delta}^{\rm (h)}_{\varsigma\hat{x}}$
and $\bar{\Delta}^{\rm (h)}_{\varsigma\hat{y}}$ satisfy following three self-consistent equations,
%\begin{widetext}
\begin{subequations}\label{SCE1}
\begin{eqnarray}
{1\over Z^{\rm (h)}_{\varsigma{\rm F}}} &=& 1+{1\over N^{2}}\sum_{{\bf{p}}{\bf{p}'}{\nu}}(-1)^{\nu+1}
{\Omega}^{(\rm{h})}_{\varsigma{\bf{p}}{\bf{p}'}{{\bf k}_{\rm A}}}\left ( {F^{(\rm{h})}_{\varsigma1\nu}({\bf{p}},{\bf{p}}',{\bf k}_{\rm A})
\over [\omega^{(\nu)}_{\varsigma{\bf{p}}{\bf{p}}'}-E^{\rm (h)}_{\varsigma}(\bf{p}+{\bf k}_{\rm A})]^{2}}
+{F^{(\rm{h})}_{\varsigma2\nu}({\bf{p}},{\bf{p}}',{\bf k}_{\rm A})\over [\omega^{(\nu)}_{\varsigma{\bf{p}}{\bf{p}}'}
+E^{\rm (h)}_{\varsigma}(\bf{p}+\bf{k}_{\rm A})]^{2}} \right ), \\
\bar{\Delta}^{\rm (h)}_{\varsigma\hat{x}} &=& {8\over N^{3}}\sum_{{\bf{p}}{\bf{p}'}{\bf{k}}{\nu}}(-1)^{\nu}Z^{\rm (h)}_{\varsigma{\rm F}}
\Omega^{(\rm{h})}_{\varsigma{\bf{p}}{\bf{p}'}{\bf{k}}}{\gamma_{{\bf k}_{x}}(\bar{\Delta}^{\rm (h)}_{\varsigma\hat{x}}
\gamma_{{\bf p}_{x}+{\bf k}_{x}}-\bar{\Delta}^{\rm (h)}_{\varsigma\hat{y}}\gamma_{{\bf p}_{y}+{\bf k}_{y}})
\over E^{\rm (h)}_{\varsigma}({\bf{p}}+{\bf{k}})}
\left( {F^{(\rm{h})}_{\varsigma1\nu}({\bf{p}},{\bf{p}}',{\bf{k}})\over
\omega^{(\nu)}_{\varsigma{\bf{p}}{\bf{p}}'}-E^{\rm (h)}_{\varsigma}(\bf{p}+\bf{k})}-{F^{(\rm{h})}_{\varsigma2\nu}({\bf{p}},{\bf{p}}',{\bf{k}})
\over\omega^{(\nu)}_{\varsigma{\bf{p}}{\bf{p}}'}+E^{\rm (h)}_{\varsigma}(\bf{p}+\bf{k})} \right ), \nonumber\\
\\
\bar{\Delta}^{\rm (h)}_{\varsigma\hat{y}} &=& {8\over N^{3}}\sum_{{\bf{p}}{\bf{p}'}{\bf{k}}{\nu}}(-1)^{\nu+1}Z^{\rm (h)}_{\varsigma{\rm F}}
\Omega^{(\rm{h})}_{\varsigma{\bf{p}}{\bf{p}'}{\bf{k}}}{\gamma_{{\bf k}_{y}}(\bar{\Delta}^{\rm (h)}_{\varsigma\hat{x}}
\gamma_{{\bf p}_{x}+{\bf k}_{x}}-\bar{\Delta}^{\rm (h)}_{\varsigma\hat{y}}\gamma_{{\bf p}_{y}+{\bf k}_{y}})
\over E^{\rm (h)}_{\varsigma}(\bf{p}+\bf{k})}
\left( {F^{(\rm{h})}_{\varsigma1\nu}({\bf{p}},{\bf{p}}',{\bf{k}})\over
\omega^{(\nu)}_{\varsigma{\bf{p}}{\bf{p}}'}-E^{\rm (h)}_{\varsigma}(\bf{p}+\bf{k})}-{F^{(\rm{h})}_{\varsigma2\nu}({\bf{p}},{\bf{p}}',{\bf{k}})
\over\omega^{(\nu)}_{\varsigma{\bf{p}}{\bf{p}}'}+E^{\rm (h)}_{\varsigma}(\bf{p}+\bf{k})} \right ),\nonumber\\
\end{eqnarray}
\end{subequations}
%\end{widetext}
respectively, where ${\bf k}_{\rm A}=[\pi,0]$. These three equations must be solved
simultaneously with following self-consistent equations,
\end{widetext}
%\begin{widetext}
\begin{subequations}\label{SCE2}
\begin{eqnarray}
\phi_{1\hat{x}}&=&{Z^{\rm (h)}_{\varsigma{\rm F}}\over N}\sum_{\bf k}\gamma_{{\bf k}_{x}}
\left ( 1-{\bar{\xi}^{(\varsigma)}_{\bf{k}}\over E^{\rm (h)}_{\varsigma}({\bf k})}{\rm{tanh}}
\left [ {1\over 2}\beta E^{\rm (h)}_{\varsigma}({\bf k})\right ] \right ), \nonumber\\
\\
\phi_{1\hat{y}}&=&{Z^{\rm (h)}_{\varsigma{\rm F}}\over N}\sum_{{\bf{k}}}\gamma_{{\bf k}_{y}}
\left ( 1-{\bar{\xi}^{(\varsigma)}_{\bf{k}}\over E^{\rm (h)}_{\varsigma}({\bf k})}{\rm{tanh}}
\left [ {1\over 2}\beta E^{\rm (h)}_{\varsigma}({\bf k})\right ] \right ),\nonumber\\
\\
\phi_{2}&=&{Z^{\rm (h)}_{\varsigma{\rm F}}\over 2N}\sum_{{\bf{k}}}\gamma_{{\bf{k}}}'
\left ( 1-{\bar{\xi}^{(\varsigma)}_{\bf{k}}\over E^{\rm (h)}_{\varsigma}({\bf k})}{\rm{tanh}}
\left [ {1\over 2}\beta E^{\rm (h)}_{\varsigma}({\bf k})\right ] \right), \nonumber\\
\\
\delta &=&{Z^{\rm (h)}_{\varsigma{\rm F}}\over 2N}\sum_{{\bf{k}}}\left ( 1-{\bar{\xi}^{(\varsigma)}_{\bf{k}}
\over E^{\rm (h)}_{\varsigma}({\bf k})}{\rm{tanh}}\left [ {1\over 2}\beta E^{\rm (h)}_{\varsigma}({\bf k})
\right ] \right ),\nonumber\\
\chi_{1\hat{x}}&=&{2\over N}\sum_{{\bf{k}}}\gamma_{{\bf k}_{x}}{B^{(\varsigma)}_{\bf{k}}
\over 2\omega^{(\varsigma)}_{\bf{k}}}\rm{coth}\left [ {1\over 2}\beta\omega^{(\varsigma)}_{\bf{k}}\right ],\\
\chi_{1\hat{y}}&=&{2\over N}\sum_{\bf k}\gamma_{{\bf k}_{y}}{B^{(\varsigma)}_{\bf{k}}
\over 2\omega^{(\varsigma)}_{\bf{k}}}\rm{coth}\left [ {1\over 2}\beta\omega^{(\varsigma)}_{\bf{k}}\right ],\\
\chi_{2}&=&{1\over N}\sum_{{\bf{k}}}\gamma_{{\bf{k}}}'{B^{(\varsigma)}_{\bf{k}}\over
2\omega^{(\varsigma)}_{\bf{k}}}\rm{coth}\left [ {1\over 2}\beta\omega^{(\varsigma)}_{\bf{k}}\right ],\\
C_{1\hat{x}}&=&{4\over N}\sum_{{\bf{k}}}\gamma^{2}_{{\bf k}_{x}}{B^{(\varsigma)}_{\bf{k}}
\over 2\omega^{(\varsigma)}_{\bf{k}}}\rm{coth}\left [ {1\over 2}\beta\omega^{(\varsigma)}_{\bf{k}}\right ],\\
C_{1\hat{y}}&=&{4\over N}\sum_{{\bf{k}}}\gamma^{2}_{{\bf k}_{y}}{B^{(\varsigma)}_{\bf{k}}
\over 2\omega^{(\varsigma)}_{\bf{k}}}\rm{coth}\left [ {1\over 2}\beta\omega^{(\varsigma)}_{\bf{k}}\right ],
\end{eqnarray}
\begin{eqnarray}
C_{1\hat{x}\hat{y}}&=&{2\over N}\sum_{{\bf{k}}}\gamma_{{\bf k}_{x}}
\gamma_{{\bf k}_{y}}{B^{(\varsigma)}_{\bf{k}}\over 2\omega^{(\varsigma)}_{\bf{k}}}\rm{coth}\left [
{1\over 2}\beta\omega^{(\varsigma)}_{\bf{k}}\right ],\\
C_{2}&=&{1\over N}\sum_{{\bf{k}}}\gamma^{2}_{{\bf{k}}'}{B^{(\varsigma)}_{\bf{k}}\over 2\omega^{(\varsigma)}_{\bf{k}}}
\rm{coth}\left [ {1\over 2}\beta\omega^{(\varsigma)}_{\bf{k}}\right],\\
C_{3\hat{x}}&=&{2\over N}\sum_{{\bf{k}}}\gamma_{{\bf k}_{x}}\gamma_{{\bf{k}}'}
{B^{(\varsigma)}_{\bf{k}}\over 2\omega^{(\varsigma)}_{\bf{k}}}\rm{coth}\left [ {1\over 2}\beta\omega^{(\varsigma)}_{\bf{k}}\right],\\
C_{3\hat{y}}&=&{2\over N}\sum_{{\bf{k}}}\gamma_{{\bf k}_{y}}\gamma_{{\bf{k}}'}
{B^{(\varsigma)}_{\bf{k}}\over 2\omega^{(\varsigma)}_{\bf{k}}}\rm{coth}\left [ {1\over 2}\beta\omega^{(\varsigma)}_{\bf{k}}\right],\\
{1\over 2}&=&{1\over N}\sum_{{\bf{k}}}{B^{(\varsigma)}_{\bf{k}}\over 2\omega^{(\varsigma)}_{\bf{k}}}\rm{coth}\left [
{1\over 2}\beta\omega^{(\varsigma)}_{\bf{k}}\right],  \\
\chi^{\rm z}_{1\hat{x}}&=&{2\over N}\sum_{{\bf{k}}}\gamma_{{\bf k}_{x}}
{B^{(\varsigma)}_{{\rm z}{\bf k}}\over 2\omega^{(\varsigma)}_{{\rm z}{\bf k}}}\rm{coth}\left [ {1\over 2}\beta
\omega^{(\varsigma)}_{{\rm z}{\bf k}} \right],  \\
\chi^{\rm z}_{1\hat{y}}&=&{2\over N}\sum_{{\bf{k}}}\gamma_{{\bf k}_{y}}
{B^{(\varsigma)}_{{\rm z}{\bf k}}\over 2\omega^{(\varsigma)}_{{\rm z}{\bf k}}}\rm{coth}\left [ {1\over 2}\beta
\omega^{(\varsigma)}_{{\rm z}{\bf k}} \right],\\
\chi^{\rm z}_{2}&=&{1\over N}\sum_{{\bf{k}}}\gamma_{{\bf{k}}'}{B^{(\varsigma)}_{{\rm z}{\bf k}}\over
2\omega^{(\varsigma)}_{{\rm z}{\bf k}}}\rm{coth}\left [ {1\over 2}\beta\omega^{(\varsigma)}_{{\rm z}{\bf k}}\right],\\
C^{\rm z}_{1\hat{x}}&=&{4\over N}\sum_{{\bf{k}}}\gamma^{2}_{{\bf k}_{x}}
{B^{(\varsigma)}_{{\rm z}{\bf k}}\over 2\omega^{(\varsigma)}_{{\rm z}{\bf k}}}\rm{coth}\left [ {1\over 2}\beta
\omega^{(\varsigma)}_{{\rm z}{\bf k}}\right],  \\
C^{\rm z}_{1\hat{y}}&=&{4\over N}\sum_{{\bf{k}}}\gamma^{2}_{{\bf k}_{y}}
{B^{(\varsigma)}_{{\rm z}{\bf k}}\over 2\omega^{(\varsigma)}_{{\rm z}{\bf k}}}\rm{coth}\left [ {1\over 2}\beta
\omega^{(\varsigma)}_{{\rm z}{\bf k}}\right],
\end{eqnarray}
\begin{eqnarray}
C^{\rm z}_{1\hat{x}\hat{y}}&=&{2\over N}\sum_{{\bf{k}}}\gamma_{{\bf k}_{x}}
\gamma_{{\bf k}_{y}}{B^{(\varsigma)}_{{\rm z}{\bf k}}\over 2\omega^{(\varsigma)}_{{\rm z}{\bf k}}}\rm{coth}
\left [ {1\over 2}\beta\omega^{(\varsigma)}_{{\rm z}{\bf k}}\right],  \\
C^{\rm z}_{3\hat{x}}&=&{2\over N}\sum_{{\bf{k}}}\gamma_{{\bf k}_{x}}\gamma_{{\bf{k}}'}
{B^{(\varsigma)}_{{\rm z}{\bf k}}\over 2\omega^{(\varsigma)}_{{\rm z}{\bf k}}}\rm{coth}\left [ {1\over 2}\beta
\omega^{(\varsigma)}_{{\rm z}{\bf k}} \right],  \\
C^{\rm z}_{3\hat{y}}&=&{2\over N}\sum_{{\bf{k}}}\gamma_{{\bf k}_{y}}\gamma_{{\bf{k}}'}
{B^{(\varsigma)}_{{\rm z}{\bf k}}\over 2\omega^{(\varsigma)}_{{\rm z}{\bf k}}}\rm{coth}\left [ {1\over 2}\beta
\omega^{(\varsigma)}_{{\rm z}{\bf k}} \right],~~~~~
\end{eqnarray}
\end{subequations}
then all the above order parameters, the decoupling parameter $\alpha$, and the charge-carrier chemical
potential $\mu_{\rm h}$ are determined by the self-consistent calculation without using any adjustable
parameters.

\begin{figure}[h!]
\centering
\includegraphics[scale=0.95]{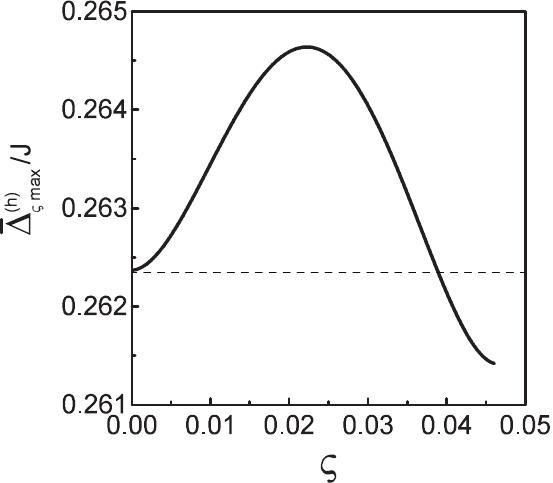}
\caption{$\bar{\Delta}^{\rm (h)}_{\varsigma{\rm max}}$ as a function of the strength of the electronic
nematicity at the optimal doping $\delta=0.15$ for $t/J=3$ and $t'/t=1/3$.
\label{pair-gap-parameter-nematicity}}
\end{figure}

The above equations (\ref{SCE1}) and (A19) have been calculated self-consistently, and the result
of the maximal charge-carrier pair gap parameter
$\bar{\Delta}^{\rm (h)}_{\varsigma{\rm max}}=\bar{\Delta}^{\rm (h)}_{\varsigma}({\bf k})|_{{\bf k}=[0,\pi]}$
in Eq. (\ref{CCPGF}) as a function of the strength of the electronic nematicity is plotted in
Fig. \ref{pair-gap-parameter-nematicity}, where with the increase of the strength of the electronic
nematicity, $\bar{\Delta}^{\rm (h)}_{\varsigma{\rm max}}$ is raised gradually in the weak strength region,
and reaches the maximum around the optimal strength of the electronic nematicity $\varsigma\approx 0.022$.
However, with the further increase of the electronic nematicity,
$\bar{\Delta}^{\rm (h)}_{\varsigma{\rm max}}$ then turns into a monotonically decrease in the
strong strength region.

\subsection{Charge-carrier pair transition temperature}\label{CCPTT}

The evolution of the charge-carrier pair transition temperature $T^{\rm (pair)}_{\rm c}$ with the doping
concentration or the strength of the electronic nematicity can be evaluated self-consistently from the
above self-consistent equations (\ref{SCE1}) and (A19) at the condition of the charge-carrier
pair gap parameter $\bar{\Delta}_{\varsigma}^{\rm (h)}=0$ [then
$\bar{\Delta}_{\varsigma\hat{x}}^{\rm (h)}=0$ and $\bar{\Delta}_{\varsigma\hat{y}}^{\rm (h)}=0$]. We
\cite{Feng15a} have shown that this $T^{\rm (pair)}_{\rm c}$ is the exactly same as that obtained from
the corresponding electron pairing state at the condition of the electron pair gap parameter
$\bar{\Delta}^{(\varsigma)}=0$, and will return to this discussion of $T^{\rm (pair)}_{\rm c}$
towards subsection \ref{EPTT} of this Appendix. The above results thus also show that (i) the mechanism
of the formation of the charge-carrier pairs is purely electronic without phonons; (ii) the mechanism
indicates that the strong correlation favors the formation of the charge-carrier pairs
(then the electron pairs), since the main ingredient is identified into a charge-carrier pairing
mechanism not involving the phonon, the external degree of freedom, but the internal spin degree of
freedom of electron; (iii) the charge-carrier pairing state is controlled by both the charge-carrier
pair gap $\bar{\Delta}_{\varsigma}^{\rm (h)}({\bf k})$ and charge-carrier quasiparticle coherent
weight $Z^{\rm (h)}_{\varsigma{\rm F}}$.

\subsection{Full charge-spin recombination}\label{FCSRS}

For the discussions of the exotic features of the electronic structure of cuprate superconductors in the
SC-state with coexisting electronic nematicity, we need to derive the full electron diagonal and
off-diagonal propagators $G_{\varsigma}({\bf k},\omega)$ and $\Im^{\dagger}_{\varsigma}({\bf k},\omega)$
in Eq. (\ref{EGF}) of the main text. In the previous studies for the case without rotation
symmetry-breaking, we \cite{Feng15a} have developed a full charge-spin recombination scheme, where a
charge carrier and a localized spin are fully recombined into a constrained electron. In particular,
within this full charge-spin recombination scheme, it has been realized that the coupling form between
the electrons and spin excitations is the same as that between the charge carriers and spin excitations,
which therefore indicates that the form of the self-consistent equations satisfied by the full electron
diagonal and off-diagonal propagators is the same as the form satisfied by the full charge-carrier
diagonal and off-diagonal propagators. Following these previous discussions \cite{Feng15a}, we can
perform a full charge-spin recombination in the present case with broken rotation symmetry in which the
full charge-carrier diagonal and off-diagonal propagators $g_{\varsigma}({\bf k},\omega)$ and
$\Gamma^{\dagger}_{\varsigma}({\bf k},\omega)$ in Eq. (\ref{CCSCES}) are replaced by the full electron
diagonal and off-diagonal propagators $G_{\varsigma}({\bf k},\omega)$ and
$\Im^{\dagger}_{\varsigma}({\bf k},\omega)$, respectively, and then the self-consistent equations
satisfied by the full electron diagonal and off-diagonal propagators of the $t$-$J$ model
(\ref{tJ-model}) in the SC-state with coexisting electronic nematicity can be obtained explicitly as,
\begin{subequations}\label{ESCES}
\begin{eqnarray}
G_{\varsigma}({\bf k},\omega)&=&G^{(0)}_{\varsigma}({\bf k},\omega)+G^{(0)}_{\varsigma}({\bf k},\omega)
[\Sigma^{(\varsigma)}_{\rm ph}({\bf k},\omega)G_{\varsigma}({\bf k},\omega)\nonumber\\
&-&\Sigma^{(\varsigma)}_{\rm pp}({\bf k},\omega)\Im^{\dagger}_{\varsigma}({\bf k},\omega)],~
\label{EDGF} \\
\Im^{\dagger}_{\varsigma}({\bf k},\omega)&=&G^{(0)}_{\varsigma}({\bf k},-\omega)
[\Sigma^{(\varsigma)}_{\rm ph}({\bf k},-\omega)\Im^{\dagger}_{\varsigma}({\bf k},\omega)\nonumber\\
&+&\Sigma^{(\varsigma)}_{\rm pp}({\bf k},\omega)G_{\varsigma}({\bf k},\omega)], \label{EODGF}
\end{eqnarray}
\end{subequations}
where $G^{(0)}_{\varsigma}({\bf k},\omega)$ is the electron diagonal propagator of the $t$-$J$ model
(\ref{tJ-model}) in the tight-binding approximation, and can be expressed explicitly as,
\begin{eqnarray}\label{MFEGF}
G^{(0)}_{\varsigma}({\bf k},\omega)&=&{1\over \omega-\varepsilon^{(\varsigma)}_{\bf k}},
\end{eqnarray}
while the electron normal self-energy $\Sigma^{(\varsigma)}_{\rm ph}({\bf k},\omega)$ in the
particle-hole channel and electron anomalous self-energy
$\Sigma^{(\varsigma)}_{\rm pp}({\bf k},\omega)$ in the particle-particle channel can be obtained
directly from the corresponding parts of the charge-carrier normal self-energy
$\Sigma^{(\rm h)}_{\varsigma{\rm ph}}({\bf k},\omega)$ and charge-carrier anomalous self-energy
$\Sigma^{(\rm h)}_{\varsigma{\rm pp}}({\bf k},\omega)$ in Eq. (\ref{CCSE}) by the replacement of
the full charge-carrier diagonal and off-diagonal propagators
$g_{\varsigma}({\bf k},\omega)$ and $\Gamma^{\dagger}_{\varsigma}({\bf k},\omega)$ with the
corresponding full electron diagonal and off-diagonal propagators $G_{\varsigma}({\bf k},\omega)$
and $\Im^{\dagger}_{\varsigma}({\bf k},\omega)$ as,
%\begin{widetext}
\begin{subequations}\label{ESE}
\begin{eqnarray}
\Sigma^{(\varsigma)}_{\rm ph}({\bf k},i\omega_{n})&=&{1\over N^{2}}\sum_{{\bf p},{\bf p}'}
[\Lambda^{(\varsigma)}_{{\bf p}+{\bf p}'+{\bf k}}]^{2}\nonumber\\
&\times& {1\over\beta}\sum_{ip_{m}}
G_{\varsigma}({{\bf p}+{\bf k}},ip_{m}+i\omega_{n})\Pi_{\varsigma}({\bf p},{\bf p}',ip_{m}),\nonumber\\
\\
\Sigma^{(\varsigma)}_{\rm pp}({\bf k},i\omega_{n})&=&{1\over N^{2}}\sum_{{\bf p},{\bf p}'}
[\Lambda^{(\varsigma)}_{{\bf p}+{\bf p}'+{\bf k}}]^{2}\nonumber\\
&\times& {1\over \beta}\sum_{ip_{m}}
\Im^{\dagger}_{\varsigma}({\bf p}+{\bf k},ip_{m}+i\omega_{n})\Pi_{\varsigma}({\bf p},{\bf p}',ip_{m}),
\nonumber\\
\end{eqnarray}
\end{subequations}
%\end{widetext}
respectively, and then the corresponding full electron diagonal and off-diagonal propagators now can be
obtained from Eq. (\ref{ESCES}) as quoted in Eq. (\ref{EGF}) of the main text. The above electron normal
self-energy $\Sigma^{(\varsigma)}_{\rm ph}({\bf k},\omega)$ describes the electron quasiparticle
coherence, and therefore gives rise to a main contribution to the energy and lifetime renormalization of
the electrons, while the electron anomalous self-energy $\Sigma^{(\varsigma)}_{\rm pp}({\bf k},\omega)$
is defined as the momentum and energy dependence of the electron pair gap,
$\Sigma^{(\varsigma)}_{\rm pp}({\bf k},\omega)=\bar{\Delta}^{(\varsigma)}({\bf k},\omega)=
\bar{\Delta}^{(\varsigma)}_{\hat{x}}({\bf k},\omega)-\bar{\Delta}^{(\varsigma)}_{\hat{y}}({\bf k},\omega)$,
and therefore is corresponding to the energy for breaking an electron pair.

In analogy to the calculation of the charge-carrier normal and anomalous self-energies in subsection
\ref{Charge-carrier-self-energy}, we can also derive explicitly the electron normal and anomalous
self-energies. Firstly, the electron normal self-energy $\Sigma^{(\varsigma)}_{\rm ph}({\bf k},\omega)$
can be separated into its symmetric and antisymmetric parts as:
$\Sigma^{(\varsigma)}_{\rm ph}({\bf k},\omega)=\Sigma^{(\varsigma)}_{\rm phe}({\bf k},\omega)
+\omega\Sigma^{(\varsigma)}_{\rm pho}({\bf k},\omega)$, with the corresponding symmetric part
$\Sigma^{(\varsigma)}_{\rm phe}({\bf k},\omega)$ and antisymmetric part
$\Sigma^{(\varsigma)}_{\rm pho}({\bf k},\omega)$ that are an even function of energy. In particular,
this antisymmetric part $\Sigma^{(\varsigma)}_{\rm pho}({\bf k},\omega)$ is defined as the electron
quasiparticle coherent weight as:
$Z^{(\varsigma)-1}_{\rm F}({\bf k},\omega)=1-{\rm Re}\Sigma^{(\varsigma)}_{\rm pho}({\bf k},\omega)$.
In an interacting electron system, everything happens at around EFS. As a case for low-energy close to
EFS, the electron pair gap and electron quasiparticle coherent weight therefore can be discussed in the
static-limit approximation,
\begin{subequations}
\begin{eqnarray}
\bar{\Delta}^{(\varsigma)}({\bf k})&=&{1\over 2}[\bar{\Delta}^{(\varsigma)}_{\hat{x}}{\rm cos}k_{x}
-\bar{\Delta}^{(\varsigma)}_{\hat{y}}{\rm cos}k_{y}],\label{EPGF}\\
{1\over Z^{(\varsigma)}_{\rm F}}&=&1-{\rm Re}
\Sigma^{(\varsigma)}_{\rm pho}({\bf k},\omega=0)\mid_{{\bf k}=[\pi,0]}, ~~~
\label{EQCW}
\end{eqnarray}
\end{subequations}
where the wave vector ${\bf k}$ in $Z^{(\varsigma)}_{\rm F}({\bf k})$ has been chosen as ${\bf k}=[\pi,0]$
just as it has been done in the ARPES experiments \cite{Ding01,DLFeng00}. Moreover, as in the case for the
charge-carrier pair gap in Eq. (\ref{CCPGF-1}), the electron pair gap in Eq. (\ref{EPGF}) can be also
expressed explicitly as,
\begin{eqnarray}
\bar{\Delta}^{(\varsigma)}({\bf k})=\bar{\Delta}^{(\varsigma)}_{\rm d}\gamma^{\rm (d)}_{\bf k}
+\bar{\Delta}^{(\varsigma)}_{\rm s}\gamma^{\rm (s)}_{\bf k},\label{EPGF-1}
\end{eqnarray}
with the d-wave component of the electron pair gap parameter $\bar{\Delta}^{(\varsigma)}_{\rm d}
=(\bar{\Delta}^{(\varsigma)}_{\hat{x}}+\bar{\Delta}^{(\varsigma)}_{\hat{y}})/2$ and the s-wave component
$\bar{\Delta}^{(\varsigma)}_{\rm s}=(\bar{\Delta}^{(\varsigma)}_{\hat{x}}-
\bar{\Delta}^{(\varsigma)}_{\hat{y}})/2$.

With the above static-limit approximation for the electron pair gap $\bar{\Delta}^{(\varsigma)}({\bf k})$
and electron quasiparticle coherent weight $Z^{(\varsigma)}_{\rm F}$, the renormalized electron diagonal
and off-diagonal propagators now can be derived directly from Eq. (\ref{ESCES}) as,
\begin{subequations}\label{MF-EGFS}
\begin{eqnarray}
G^{\rm (RMF)}_{\varsigma}({\bf k},\omega)&=&Z^{(\varsigma)}_{\rm F}\left ({U^{(\varsigma)2}_{\bf k}\over
\omega-E^{(\varsigma)}_{\bf k}}+{V^{(\varsigma)2}_{\bf k}\over\omega +E^{(\varsigma)}_{\bf k}} \right ), ~~~~~~\\
\Im^{{\rm (RMF)}\dagger}_{\varsigma}({\bf k},\omega)&=&-Z^{(\varsigma)}_{\rm F}
{\bar{\Delta}^{(\varsigma)}_{\rm Z}({\bf k})\over 2E^{(\varsigma)}_{\bf k}}\left ({1\over\omega
-E^{(\varsigma)}_{\bf k}}\right .\nonumber\\
&-&\left . {1\over\omega+E^{(\varsigma)}_{\bf k}}\right ),
\end{eqnarray}
\end{subequations}
where the renormalized electron orthorhombic energy dispersion
$\bar{\varepsilon}^{(\varsigma)}_{\bf k}=Z^{(\varsigma)}_{\rm F}\varepsilon^{(\varsigma)}_{\bf k}$, the
renormalized electron pair gap
$\bar{\Delta}^{(\varsigma)}_{\rm Z}({\bf k})=Z^{(\varsigma)}_{\rm F}\bar{\Delta}^{(\varsigma)}({\bf k})$,
the SC quasiparticle energy spectrum $E^{(\varsigma)}_{\bf k}=\sqrt {\bar{\varepsilon}^{(\varsigma)2}_{\bf k}
+|\bar{\Delta}^{(\varsigma)}_{\rm Z}({\bf k})|^{2}}$, and the SC quasiparticle coherence factors
\begin{subequations}\label{EBCSCF}
\begin{eqnarray}
U^{(\varsigma)2}_{\bf k}&=&{1\over 2}\left ( 1+{\bar{\varepsilon}^{(\varsigma)}_{\bf k}\over
E^{(\varsigma)}_{\bf k}}\right ), \\
V^{(\varsigma)2}_{\bf k}&=&{1\over 2}\left ( 1-{\bar{\varepsilon}^{(\varsigma)}_{\bf k}\over
E^{(\varsigma)}_{\bf k}}\right ),
\end{eqnarray}
\end{subequations}
with the constraint $U^{(\varsigma)2}_{\bf k}+V^{(\varsigma)2}_{\bf k}=1$ for any wave vector ${\bf k}$. The
above result in Eq. (\ref{EPGF-1}) therefore show that the symmetry of superconductivity with coexisting
electronic nematicity is modified from the ordinary d-wave electron pairing to the d+s wave
\cite{Kitatani17,Maier14}. Moreover, the results in Eqs. (\ref{MF-EGFS}) and (\ref{EBCSCF}) are the standard
Bardeen-Cooper-Schrieffer expressions for an electron pair state \cite{Schrieffer64}, although the
electron pairing mechanism is driven by the kinetic energy by the exchange of a strongly dispersive spin
excitation.

Substituting these renormalized electron diagonal and off-diagonal propagators
in Eq. (\ref{MF-EGFS}) and MF spin propagator in Eq. (\ref{MFSGF}) into Eqs. (\ref{ESE}), the electron
normal and anomalous self-energies can be obtained explicitly as,
\begin{widetext}
\begin{subequations}\label{ESE-1}
\begin{eqnarray}
{\Sigma}^{(\varsigma)}_{\rm ph}({\bf{k}},{\omega})&=&\frac{1}{N^{2}}\sum_{{\bf{p}}{\bf{p}'}{\nu}}(-1)^{\nu+1}
{\Omega}^{(\varsigma)}_{{\bf{p}}{\bf{p}'}{\bf{k}}}\left [ U^{(\varsigma)2}_{\bf{p}+\bf{k}}\left (
\frac{F^{(\varsigma)}_{1\nu}({\bf p},{\bf p}',{\bf k})}{\omega+\omega^{(\nu)}_{\varsigma{\bf{p}}{\bf{p}}'}
-E^{(\varsigma)}_{\bf{p}+\bf{k}}}-\frac{F^{(\varsigma)}_{2\nu}({\bf p},{\bf p}',{\bf k})}{\omega
-\omega^{(\nu)}_{\varsigma{\bf{p}}{\bf{p}}'}-E^{(\varsigma)}_{\bf{p}+\bf{k}}}\right)\right. \nonumber\\
&+&\left . V^{(\varsigma)2}_{\bf{p}+\bf{k}} \left (\frac{F^{(\varsigma)}_{1\nu}({\bf p},{\bf p}',{\bf k})}{\omega- \omega^{(\nu)}_{\varsigma{\bf{p}}{\bf{p}}'}+E^{(\varsigma)}_{\bf{p}+\bf{k}}}
-\frac{F^{(\varsigma)}_{2\nu}({\bf p},{\bf p}',{\bf k})}
{\omega+\omega^{(\nu)}_{\varsigma{\bf{p}}{\bf{p}}'}+E^{(\varsigma)}_{\bf{p}+\bf{k}}}  \right )  \right ],
\label{ph-ESE}\\
{\Sigma}^{(\varsigma)}_{\rm pp}({\bf{k}},{\omega})&=&\frac{1}{N^{2}}\sum_{{\bf{p}}{\bf{p}'}{\nu}}(-1)^{\nu}
{\Omega}^{(\varsigma)}_{{\bf{p}}{\bf{p}'}{\bf{k}}}\frac{\bar{\Delta}_{\varsigma{\rm{Z}}}({\bf{p}}+{\bf{k}})}
{2E^{(\varsigma)}_{{\bf{p}}+\bf{k}}}\left [ \left ( \frac{F^{(\varsigma)}_{1\nu}({\bf p},{\bf p}',{\bf k})} {\omega+\omega^{(\nu)}_{\varsigma{\bf{p}}{\bf{p}}'}-E^{(\varsigma)}_{\bf{p}+\bf{k}}}
-\frac{F^{(\varsigma)}_{2\nu}({\bf p},{\bf p}',{\bf k})}{\omega-\omega^{(\nu)}_{\varsigma{\bf{p}}{\bf{p}}'}
-E^{(\varsigma)}_{\bf{p}+\bf{k}}} \right)\right. \nonumber\\
&-&\left. \left ( \frac{F^{(\varsigma)}_{1\nu}({\bf p},{\bf p}',{\bf k})} {\omega-\omega^{(\nu)}_{\varsigma{\bf{p}}{\bf{p}}'}+E^{(\varsigma)}_{\bf{p}+\bf{k}}}
-\frac{F^{(\varsigma)}_{2\nu}({\bf p},{\bf p}',{\bf{k}})}{\omega+\omega^{(\nu)}_{\varsigma{\bf{p}}{\bf{p}}'}
+E^{(\varsigma)}_{\bf{p}+\bf{k}}}  \right )  \right ], \label{pp-ESE}
\end{eqnarray}
\end{subequations}
%\end{widetext}
respectively, where $\nu=1,2$, ${\Omega}^{(\varsigma)}_{{\bf{p}}{\bf{p}'}{\bf{k}}}=Z^{(\varsigma)}_{\rm{F}}[\Lambda^{(\varsigma)}_{{\bf{p}}
+{\bf{p}'}+{\bf{k}}}]^{2}B^{(\varsigma)}_{\bf{p}'}B^{(\varsigma)}_{\bf{p}+\bf{p}'}
/(4\omega^{(\varsigma)}_{\bf{p}'}\omega^{(\varsigma)}_{\bf{p}+\bf{p}'})$,
and the functions,
%\begin{widetext}
\begin{subequations}
\begin{eqnarray}
F^{(\varsigma)}_{1\nu}({\bf{p}},{\bf{p}}',{\bf{k}})&=& n_{\rm{F}}(E^{(\varsigma)}_{\bf{p}+\bf{k}})\{1
+n_{B}(\omega^{(\varsigma)}_{\bf{p}'+\bf{p}})+n_{B}[(-1)^{\nu+1}\omega^{(\varsigma)}_{\bf{p}'}]\}
+n_{B}(\omega^{(\varsigma)}_{\bf{p}'+\bf{p}})n_{B}[(-1)^{\nu+1}\omega^{(\varsigma)}_{\bf{p}'}], \\
F^{(\varsigma)}_{2\nu}({\bf{p}},{\bf{p}}',{\bf{k}})&=& [1-n_{\rm{F}}(E^{(\varsigma)}_{\bf{p}+\bf{k}})]\{
1+n_{B}(\omega^{(\varsigma)}_{\bf{p}'+\bf{p}})+n_{B}[(-1)^{\nu+1}\omega^{(\varsigma)}_{\bf{p}'}] \}
+n_{B}(\omega^{(\varsigma)}_{\bf{p}'+\bf{p}})n_{B}[(-1)^{\nu+1}\omega^{(\varsigma)}_{\bf{p}'}].~~~~
\end{eqnarray}
\end{subequations}
%\end{widetext}

%\begin{widetext}
\subsection{Self-consistent equations for determination of electron order parameters}\label{ESCE}

The above electron quasiparticle coherent weight, two components of the electron pair gap parameter,
and the electron chemical potential satisfy following four self-consistent equations,
%\begin{widetext}
\begin{subequations}\label{SCE3}
\begin{eqnarray}
\frac{1}{Z^{(\varsigma)}_{\rm{F}}} &=& 1+\frac{1}{N^{2}}\sum_{{\bf{p}}{\bf{p}'}{\nu}}(-1)^{\nu+1}
\Omega^{(\varsigma)}_{{\bf{p}}{\bf{p}'}{\bf k}_{\rm A}}\left (
\frac{F^{(\varsigma)}_{1\nu}({\bf p},{\bf p}',{\bf k}_{\rm A})}{(\omega^{(\nu)}_{\varsigma{\bf{p}}{\bf{p}}'}
-E^{(\varsigma)}_{\bf{p}+{\bf k}_{\rm A}})^{2}}
+\frac{F^{(\varsigma)}_{2\nu}({\bf{p}},{\bf{p}}',{\bf k}_{\rm A})}
{(\omega^{(\nu)}_{\varsigma{\bf{p}}{\bf{p}}'}+E^{(\varsigma)}_{\bf{p}+{\bf k}_{\rm A}})^{2}} \right ), \\
\bar{\Delta}^{(\varsigma)}_{\hat{x}} &=& \frac{8}{N^{3}}\sum_{{\bf{p}}{\bf{p}'}{\bf{k}}{\nu}}(-1)^{\nu}
Z^{(\varsigma)}_{\rm{F}}\Omega^{(\varsigma)}_{{\bf{p}}{\bf{p}'}{\bf{k}}}\frac{\gamma_{{\bf k}_{x}}
(\bar{\Delta}^{(\varsigma)}_{\hat{x}}\gamma_{{\bf p}_{x}+{\bf k}_{x}}-\bar{\Delta}^{(\varsigma)}_{\hat{y}}
\gamma_{{\bf p}_{y}+{\bf k}_{y}})}{E^{(\varsigma)}_{\bf{p}+\bf{k}}}\left (
\frac{F^{(\varsigma)}_{1\nu}({\bf{p}},{\bf{p}}',{\bf{k}})}{\omega^{(\nu)}_{\varsigma{\bf{p}}{\bf{p}}'}
-E^{(\varsigma)}_{\bf{p}+\bf{k}}}-\frac{F^{(\varsigma)}_{2\nu}({\bf{p}},{\bf{p}}',{\bf{k}})}
{\omega^{(\nu)}_{\varsigma{\bf{p}}{\bf{p}}'}+E^{(\varsigma)}_{\bf{p}+\bf{k}}} \right ),~~~~~ \\
\bar{\Delta}^{(\varsigma)}_{\hat{y}} &=& \frac{8}{N^{3}}\sum_{{\bf{p}}{\bf{p}'}{\bf{k}}{\nu}}(-1)^{\nu+1}
Z^{(\varsigma)}_{\rm{F}}\Omega^{(\varsigma)}_{{\bf{p}}{\bf{p}'}{\bf{k}}}\frac{\gamma_{{\bf k}_{y}}
(\bar{\Delta}^{(\varsigma)}_{\hat{x}}\gamma_{{\bf p}_{x}+{\bf k}_{x}}-\bar{\Delta}^{(\varsigma)}_{\hat{y}}
\gamma_{{\bf p}_{y}+{\bf k}_{y}})}{E^{(\varsigma)}_{\bf{p}+\bf{k}}} \left (
\frac{F^{(\varsigma)}_{1\nu}({\bf{p}},{\bf{p}}',{\bf{k}})}{\omega^{(\nu)}_{\varsigma{\bf{p}}{\bf{p}}'}
-E^{(\varsigma)}_{\bf{p}+\bf{k}}}-\frac{F^{(\varsigma)}_{2\nu}({\bf{p}},{\bf{p}}',{\bf{k}})}
{\omega^{(\nu)}_{\varsigma{\bf{p}}{\bf{p}}'}+E^{(\varsigma)}_{\bf{p}+\bf{k}}} \right ),~~~~~ \\
1-\delta &=&\frac{1}{2N}\sum_{{\bf{k}}}Z^{(\varsigma)}_{\rm{F}} \left(
1-\frac{\bar{\varepsilon}^{(\varsigma)}_{\bf{k}}}{E^{(\varsigma)}_{{\bf{k}}}}{\rm{tanh}}
\left[ \frac{1}{2}\beta{E^{(\varsigma)}_{\bf{k}}} \right ] \right ),\label{SCE-EFS}
\end{eqnarray}
\end{subequations}
\end{widetext}
where ${\bf k}_{\rm A}=[\pi,0]$. In the same calculation condition as in the evaluation of the
self-consistent equations (\ref{SCE1}) and (A19), the above self-consistent equations
(\ref{SCE3}) have been also solved simultaneously, and then the electron quasiparticle coherent
weight, two components of the electron pair gap parameter, and electron chemical potential are
obtained without using any adjustable parameters. In particular, the dome-like shape
nematic-order strength dependence of $\bar{\Delta}^{\rm (h)}_{\varsigma{\rm max}}$ shown in
Fig. \ref{pair-gap-parameter-nematicity} therefore also leads to the same dome-like shape
nematic-order strength dependence of $\bar{\Delta}^{(\varsigma)}_{\rm max}$ in Eq. (\ref{EPGF}).
Moreover, it should be emphasized that the self-consistent equation in Eq. (\ref{SCE-EFS})
guarantees that the EFS contour in the presence of the electronic nematicity still satisfy
Luttinger's theorem \cite{Luttinger60}, i.e., the effective area of the EFS contour contains
$1-\delta$ electrons.

\subsection{Electron pair transition temperature}\label{EPTT}

Concomitantly, the evolution of the electron pair transition temperature $T_{\rm c}$ with the
doping or the strength of the electronic nematicity can be obtained self-consistently from the
above self-consistent equations in Eq. (\ref{SCE3}) at the condition of the electron pair gap
parameter $\bar{\Delta}^{(\varsigma)}=0$ [then $\bar{\Delta}^{(\varsigma)}_{\hat{x}}=0$ and
$\bar{\Delta}^{(\varsigma)}_{\hat{y}}=0$].

In our previous study \cite{Feng15a}, it has been demonstrated that in a given doping concentration,
the magnitude of the electron pair transition temperature $T_{\rm c}$ obtained from the electron
pairing state is the exactly same as the magnitude of the charge-carrier pair transition temperature
$T^{\rm (pair)}_{\rm c}$ obtained from the corresponding charge-carrier pairing state. Within the
framework of the kinetic-energy-driven superconductivity \cite{Feng15,Feng0306,Feng12,Feng15a}, the
effective attractive interaction between charge carriers originates in their coupling to a strongly
dispersive spin excitation, while the electron pairing interaction in the full charge-spin
recombination scheme \cite{Feng15a} is mediated by the same spin excitation, which therefore leads
to that the electron pair transition temperature $T_{\rm c}$ is identical to the charge-carrier pair
transition-temperature $T^{\rm (pair)}_{\rm c}$, and then the dome-like shape of the doping dependence
of $T_{\rm c}$ with its maximum occurring at around the optimal doping is a natural consequence of
the dome-like shape of the doping dependence of $T^{\rm (pair)}_{\rm c}$ with its maximum occurring at
around the same optimal doping.

In the SC-state with coexisting electronic nematicity, our present results therefore show that the
value of $T_{\rm c}$ obtained self-consistently from the equations (\ref{SCE3}) also is the exactly
same as the corresponding value of $T^{\rm (pair)}_{\rm c}$ obtained self-consistently in subsection
\ref{CCPTT} of this Appendix from the equations (\ref{SCE1}) and (A19) in a given doping
concentration and strength of the electronic nematicity, and then the dome-like shape nematic-order
strength dependence of the optimized $T_{\rm c}$ with its maximum appearing at around the optimal
strength of the electronic nematicity also is a natural consequence of the dome-like shape nematic-order
strength dependence of the optimized $T^{\rm (pair)}_{\rm c}$ with its maximum appearing at around the
same optimal doping.

\end{appendix}

\end{document}